\documentclass[journal]{IEEEtran}
\usepackage{amsmath,amsfonts}
\usepackage{algorithmic}
\usepackage{algorithm}
\usepackage{array}
\usepackage[caption=false,font=normalsize,labelfont=sf,textfont=sf]{subfig}
\usepackage{textcomp}
\usepackage{stfloats}
\usepackage{url}
\usepackage{verbatim}
\usepackage{graphicx}
\usepackage{cite}
\usepackage[english]{babel}

\usepackage{amsmath}
\usepackage{graphicx}
\usepackage{multirow}
\usepackage[colorlinks=true, allcolors=blue]{hyperref}

\usepackage{array}
\newcolumntype{L}[1]{>{\raggedright\let\newline\\\arraybackslash\hspace{0pt}}m{#1}}
\newcolumntype{C}[1]{>{\centering\let\newline\\\arraybackslash\hspace{0pt}}m{#1}}
\newcolumntype{R}[1]{>{\raggedleft\let\newline\\\arraybackslash\hspace{0pt}}m{#1}}

\begin{document}

\title{Affective Game Computing: A Survey}
\author{
    \IEEEauthorblockN{Georgios N. Yannakakis, \emph{IEEE Senior Member}} and David Melhart
    \\
    \IEEEauthorblockA{\textit{Institute of Digital Games, University of Malta}\\
    Msida, Malta \\
    georgios.yannakakis@um.edu.mt, david.melhart@um.edu.mt}
}



\maketitle

\begin{abstract}
This paper surveys the current state of the art in affective computing principles, methods and tools as applied to games. We review this emerging field, namely \emph{affective game computing}, through the lens of the four core phases of the affective loop: \emph{game affect elicitation}, \emph{game affect sensing}, \emph{game affect detection} and \emph{game affect adaptation}. In addition, we provide a taxonomy of terms, methods and approaches used across the four phases of the affective game loop and situate the field within this taxonomy. We continue with a comprehensive review of available affect data collection methods with regards to gaming interfaces, sensors, annotation protocols, and available corpora. The paper concludes with a discussion on the current limitations of \emph{affective game computing} and our vision for the most promising future research directions in the field. 
\end{abstract}


\section{Introduction}



\IEEEPARstart{O}{ver} a third of the Earth's population is playing games by now---with the projected number of gamers rising up to 3.3 billion by 2024 \cite{clement2022stats}. We could argue that game playing at this massive scale is probably the largest ongoing experiment of human behaviour and experience. The emotional patterns that a player goes through are deeply interwoven in the design of any game. That central role of affect interaction in this domain make games an ideal test-bed for the study of affective computing (AC) \cite{picard1995affective}. Games, however, are not merely an important domain for AC. As a matter of fact, games have shaped and advanced the AC field in numerous ways given the unique challenges they pose and opportunities they bring to affective interaction. 

Looking at digital games through the lens of the \emph{affective loop} \cite{hook2008affective}, one can only observe benefits for AC research and innovation. When it comes to emotion \emph{elicitation}, games define one of the richest forms of human-computer interaction and thus offer highly multimodal and dynamic ways to elicit affect. Moving on to affect \emph{sensing}, the availability of game engine and sensor technology brings AC researchers unprecedented opportunities for measuring manifestations of affect way beyond physiology, verbal and non-verbal communication (e.g. game analytics and in-game social activity). Affect \emph{detection} benefits from the massive gameplay corpora available in the wild e.g. over streaming services \cite{melhart2020moment}. Finally, affect \emph{adaptation} in games can be achieved via highly diverse stimuli that vary from AI-controlled expressive agents to content generators of various types \cite{liapis2014computational,liapis2018orchestrating}.

In this paper, we survey the emerging research area at the intersection of affective computing and games, namely \emph{affective game computing}. In particular, we build on the affective loop paradigm \cite{sundstrom2005exploring,hook2008affective} (see Section \ref{sec:loop}) and survey core contributions in the areas of affect \emph{elicitation}, \emph{sensing}, \emph{detection} and \emph{adaptation} in games (Sections \ref{sec:elicitation}--\ref{sec:adaptation}). We use indicative examples from both the game industry and academic research showcasing the advancements of AC through games but also the benefits AC offers to games and their development. Throughout our survey, we identify core terms, methods and approaches that we, later on, use to situate the affective game computing field as a whole (Section \ref{sec:holistic}). Moreover, our survey puts an emphasis on state-of-the-art data collection methods in games relating to interfaces, sensing devices, and annotation protocols (Section \ref{sec:dataCollection}) and the available affect corpora (Section \ref{sec:corpora}). The paper concludes with a detailed list of current limitations of the available methods and technologies (Section \ref{sec:discussion}) and outlines a number of promising future research directions for \emph{affective game computing} (Section \ref{sec:road}). We feel (and hope) that all aforementioned studies, methods, resources and tools contained in this survey paper will serve as a guide for \emph{affective game computing} researchers and will also lower the entry bar for any newcomer to this emerging research and innovation field. 

\subsection{Contributions of this Paper}

The field of affective computing in the domain of games has been studied extensively over the last 15 years. The literature is rich in this application area---as one can observe through the volume of references in this paper. Despite the variety and breadth of the studies covered, however, only a few papers have reviewed this research field in a comprehensive and detailed manner to the degree this paper does. Indicatively, an early short survey of the field focused on the relationship between emotion and games \cite{yannakakis2014emotion} and introduced the concept of the affective loop in games. The edited volume \emph{Emotion In Games} \cite{karpouzis2016emotion} provides broad coverage of several aspects of affective computing research in games but it does not survey the field comprehensively and in a systematic fashion. In \cite{yannakakis2018artificial} Yannakakis and Togelius offer an entire chapter on player modelling---the use of computational means to capture aspects of playing behaviour and experience---which touches upon some of the aspects covered here. Two more recent relevant surveys include the work of Robinson and Clore \cite{robinson2020let} who examined the use of physiological sensors in the field of human-computer interaction and Navarro et al. \cite{navarro2021biofeedback} who surveyed biofeedback interaction specifically in videogames for general entertainment.

In contrast to all the aforementioned attempts, this paper introduces the \emph{affective game computing} field and surveys it in a holistic, systematic, and comprehensive manner. In particular, the paper covers both industry and academic examples, and it offers a taxonomy of terms and methods through which we can map the critical studies in the field. Furthermore, the paper surveys existing corpora, methods and annotation protocols, provides guidelines for the newcomer in this field and discusses the next most promising research steps forward.  


\section{Affective Game Loop} \label{sec:loop}

\begin{figure}[!tb] 
    \centering
    \includegraphics[width=1\linewidth]{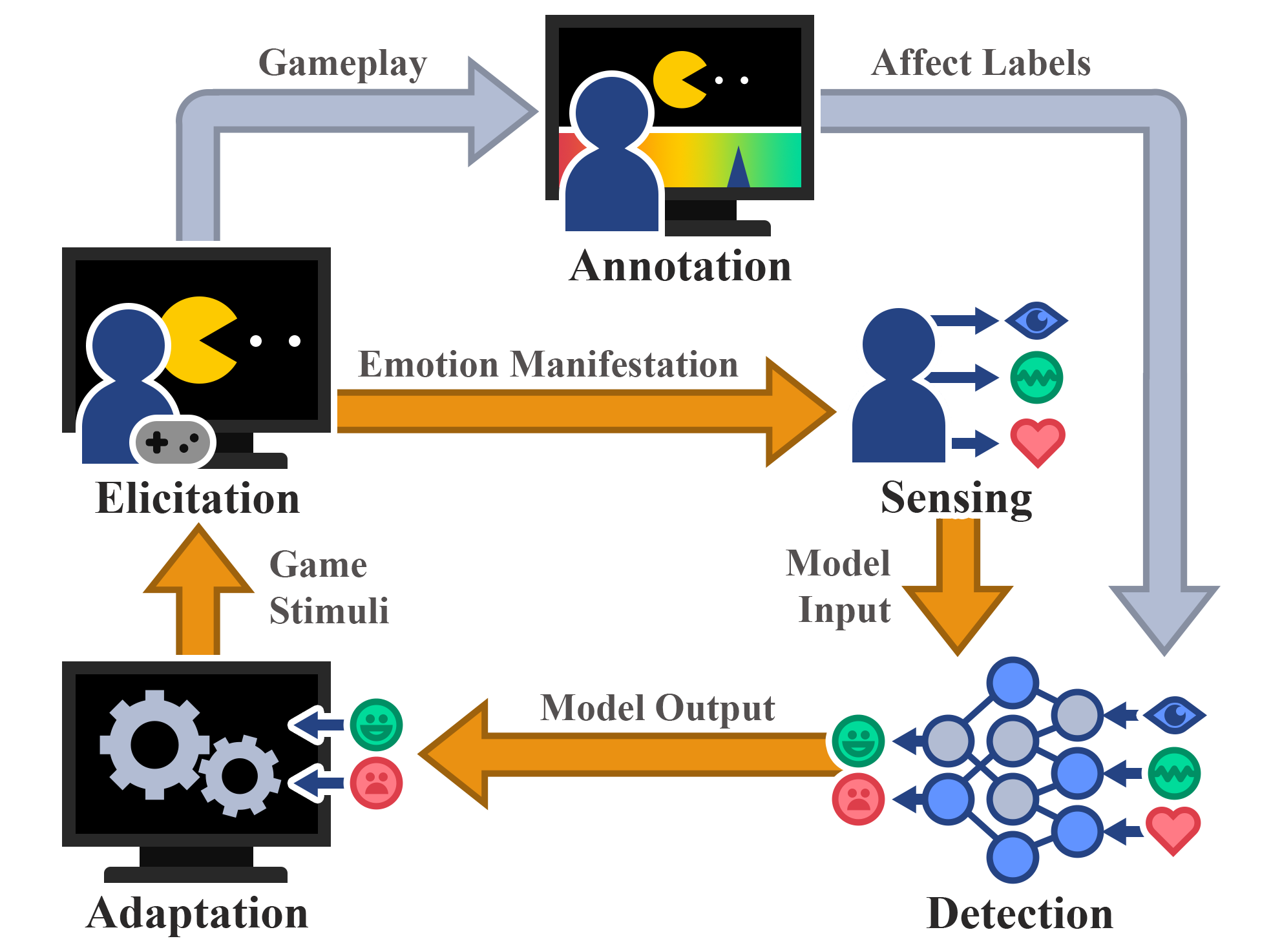}
    \caption{A high-level illustration of the affective game loop. The orange arrows depict the core steps of the loop via elicitation, sensing, adaptation, and detection. The grey arrows show how affect annotation can be integrated into the loop.}
    \label{fig:affective_loop}
\end{figure}

The affective loop first introduced by Sundström \cite{sundstrom2005exploring} and Höök \cite{hook2008affective} has become a dominant AC paradigm that is able to represent any affective interaction in a general fashion. The affective loop comprises four core sequential phases that enable an affective interaction: affect \emph{elicitation}, affect \emph{sensing}, affect \emph{detection} and affect \emph{adaptation}. When the affective loop principle is applied to games the resulting paradigm has been defined as the \emph{affective game loop} \cite{yannakakis2014emotion} (see Fig. \ref{fig:affective_loop}). Before delving into the details of the AC and games survey, in this section, we outline the key elements of each phase of the affective game loop as follows. 

\begin{enumerate}
    \item \textbf{Elicitation:} In this initial phase of the loop the game yields affective responses to players via a multitude of available affect elicitors such as game agents and game content. We detail those elicitors and survey the corresponding literature on game affect elicitation in Section \ref{sec:elicitation}.    
    \item \textbf{Sensing:} Once affect is elicited, players manifest it in numerous ways. The second phase of the affective game loop is responsible for sensing those manifestations via sensor and tracking technology as detailed in Section \ref{sec:sensing}.
    \item \textbf{Detection:} Given appropriate signals obtained from the sensed multimodal player input and human demonstrations of affect (such as player experience annotations) one can build mathematical formulations (e.g. via statistical machine learning) that are capable of inferring the annotated affect accurately based on the user signals. Details on game affect detection methods are provided in Section \ref{sec:detection}.
    \item \textbf{Adaptation:} In the last phase of the affective game loop the game is required to offer the next sequence of in-game stimuli so that the experience of the player is set within predetermined bounds. We survey methods and studies on game affect adaptation in Section \ref{sec:adaptation}.
\end{enumerate}

\section{Game Affect Elicitation}\label{sec:elicitation}

As mentioned earlier, games are equipped with a rather diverse set of stimuli that are capable of eliciting a wide spectrum of emotions in players. In this section, we provide a taxonomy of such stimuli independently of the affective states they might be able to elicit. 
Importantly, the context of the game environment, the game genre, the form of interfacing, the number of players, potential social aspects of the game, and the overall objective of the game are foundational and they impact any other in-game elicitor covered here. For our taxonomy of elicitors we largely adopt and build upon the taxonomies introduced by Yannakakis and Togelius \cite{yannakakis2018artificial}. In particular, as summarised in Table \ref{tab:elicitors} we identify three categories of affect elicitors in games namely \emph{game context}, \emph{game agent} and \emph{game content}. We discuss these categories in detail below followed by an indicative example of affect elicitation in games in Section~\ref{sec:elicitation:example}. 

\begin{figure}
    \centering
    \includegraphics[width=1\linewidth]{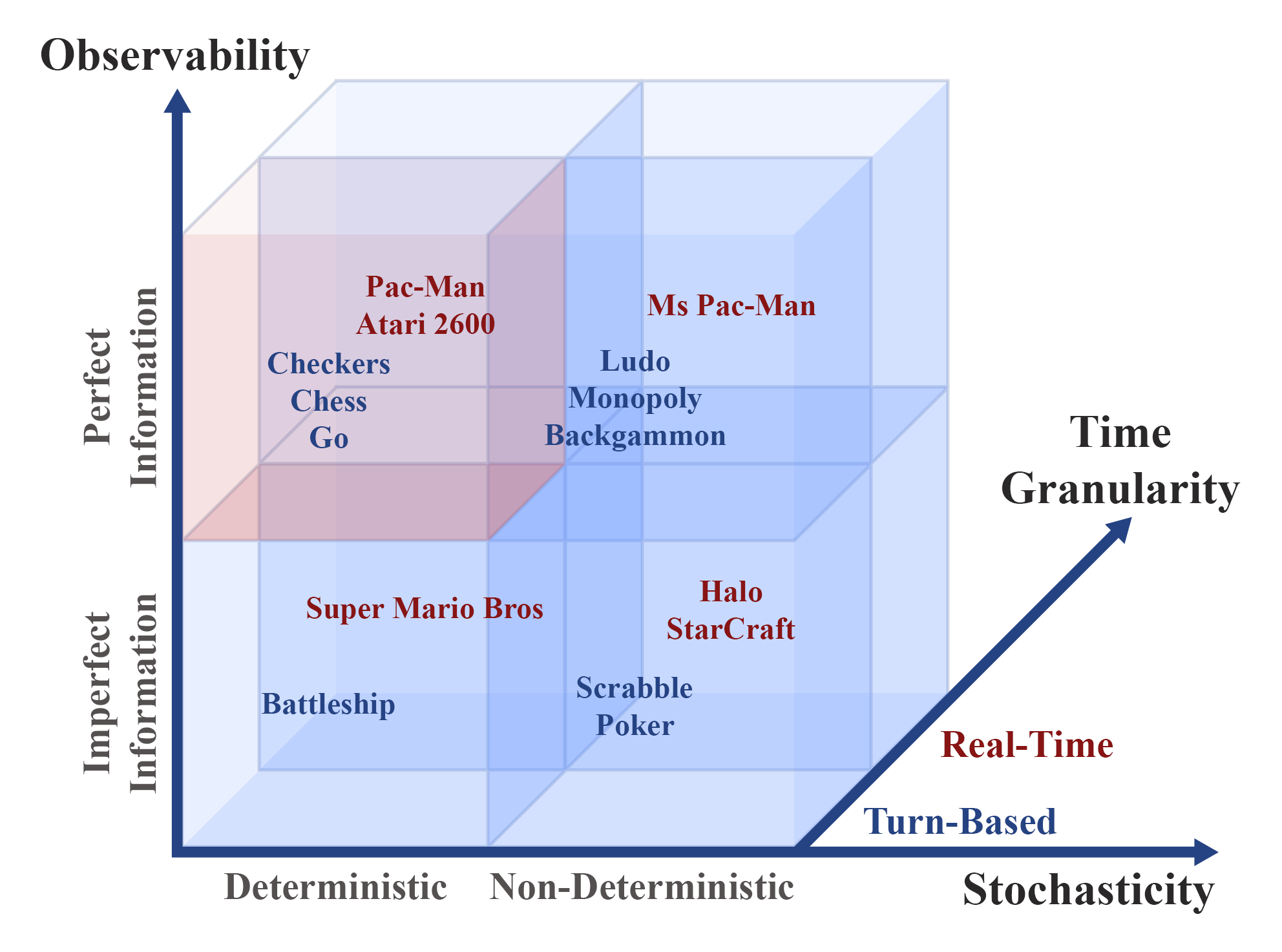}
    \caption{Characteristics of Games: Examples across the dimensions of Stochasticity, Observability, and Time Granularity. While classical algorithms such as minimax are only able to solve games in the red square, AI methods that can approximate a decision tree (e.g. Monte Carlo Tree Search) can solve games in the blue squares as well. Adopted from \cite{yannakakis2018artificial} with authors' permission.}
    \label{fig:game_characteristics}
\end{figure}

Game context refers to the game's genre, the platform used, and the game's characteristics that collectively define the momentaneous state of the game. A player can obviously affect the dynamic aspects of the game context (i.e. the game state) and vice versa, the game context can affect the gameplay and elicit affect patterns. The importance of game context is critical for player affect modelling as the context of the game needs to be considered for reliable affect detection. Simply put, any player's reactions cannot be dissociated from the stimulus that elicited them. Following the taxonomy introduced in \cite{yannakakis2018artificial} the core game characteristics that fall under the game context group include: the \emph{number of players}, the \emph{observability} of play, the \emph{stochasticity} of the game, the \emph{time granularity}, and the \emph{action space} for the player (see Fig.~\ref{fig:game_characteristics}). Under the game agent category, we fit any affect elicitor related to AI-controlled agents that might be available in the game. Examples include agent facial expression, verbal, and non-verbal agent behaviour, social agent behaviour, and individual agent behavioural patterns that can affect a player's experience. Finally, game content refers to any content type existent in a game that is not related to AI agent behaviours as those are covered in the game agent category (i.e. the virtual environment \cite{reetz2021nature}). Building on the categorisation of Liapis et al. \cite{liapis2014computational} each game is viewed as a synthesis of creative facets available which include the design of the games (i.e. rules and mechanics), the game level, and the creative ways a human plays the game (i.e. gameplay). Depending on the game available content types may include visuals, audio, narrative, and even novel control modalities \cite{canossa2020hold}. 

\begin{table*}[!tb]
\caption{A Taxonomy of Game Affect Elicitors}  
\centering
\begin{tabular}{|l||l|l|}
\hline 
\textbf{Category} & \textbf{Subcategory}  & \textbf{Elicitors} \\ \hline \hline
 & Genre & Shooter, platformer, racing, strategy, adventure, etc. \\
 & Platform & Mobile, AR, VR, console, desktop. \\
 & Number of players & single, one-and-a-half, two-player, multi-player \\
Context & Observability & Fully observable vs. partially observable game \\ 
 & Stochasticity & Deterministic vs. non-deterministic game  \\ 
 & Time granularity  & Real-time vs. continuous  \\ 
 & Action Space & Player actions available: 2 to many   \\ \hline
AI Agent &  & Navigation, expression, exploration, verbal/non-verbal interaction\\ \hline
Content &  & Visuals, Audio, Narrative, Game Design, Levels, Gameplay \cite{liapis2014computational}\\ \hline
\end{tabular}
\label{tab:elicitors}
\end{table*}

All the above-mentioned elicitors can affect the experience of play and are met in various configurations in games. It is rarely the case---as in most other domains of AC---that only one elicitor type (e.g. an image, or a sound) is active at a time. In games, instead, affect elicitors are \emph{orchestrated} \cite{liapis2018orchestrating,graja2020impact} in groups thereby offering rich and multifaceted affective interactions \cite{ravaja2016virtual,lopes2017modelling,graja2020impact,ogawa2020auditory,canossa2020hold,colombo2021psychometric}. As an example of such an orchestration process think of a scary story that is told by an expressive agent who is placed in a dungeon level with the corresponding visual and audio effects, and the appropriate virtual camera placement, and lighting. 

It is important to note that the game context is \emph{static} meaning that no aspect of it can be altered by the player or the game. As a result, the game context cannot act as a dynamic affect stimulus when the affective loop reaches its adaptation phase. One could think of games that change their genre, their mode of interaction (e.g. from VR to desktop) \cite{born2021exergaming} and the number of players but those alterations are rare. Therefore in this section, instead of surveying the existing literature with respect to all possible affect elicitors in games, we will survey only the \emph{game context} category. We will then survey AI \emph{agents} and \emph{content} in Section \ref{sec:adaptation} as these two categories include \emph{dynamic} stimuli that can be altered during the game affective interaction. 

Aspects of the game context have been predominantly included as a modality of input for affect detection in games (e.g. \cite{mcquiggan2007early,conati2009modeling,pedersen2010modeling,martinez2011mining,Shaker10AIIDE,robison2009evaluating,hazlett2006measuring,ravajapsychophysiology,mcquiggan2008modeling,martinez2010genetic} among others). In most of these studies (e.g. \cite{martinez2011mining,pedersen2010modeling,martinez2010genetic}) the game context does not refer to any of the aspects included in Table \ref{tab:elicitors} but rather to aspects of the game environment such as level features. We could argue that the game context aspects covered in Table \ref{tab:elicitors} become highly relevant for affect detection once one performs studies on general affect modelling \cite{camilleri2017towards,melhart2022again,makantasis2019pixels,makantasis2021pixels}. Very few studies have explored modalities of user input in isolation of the game context. For example, Makantasis et al. \cite{makantasis2021privileged} built models of affect based solely on physiology for the purpose of comparison against models that fused aspects of in-game video and audio. 

\subsection{Indicative Examples}\label{sec:elicitation:example}

\begin{figure}[!tb] 
    \centering
    \includegraphics[width=1\linewidth]{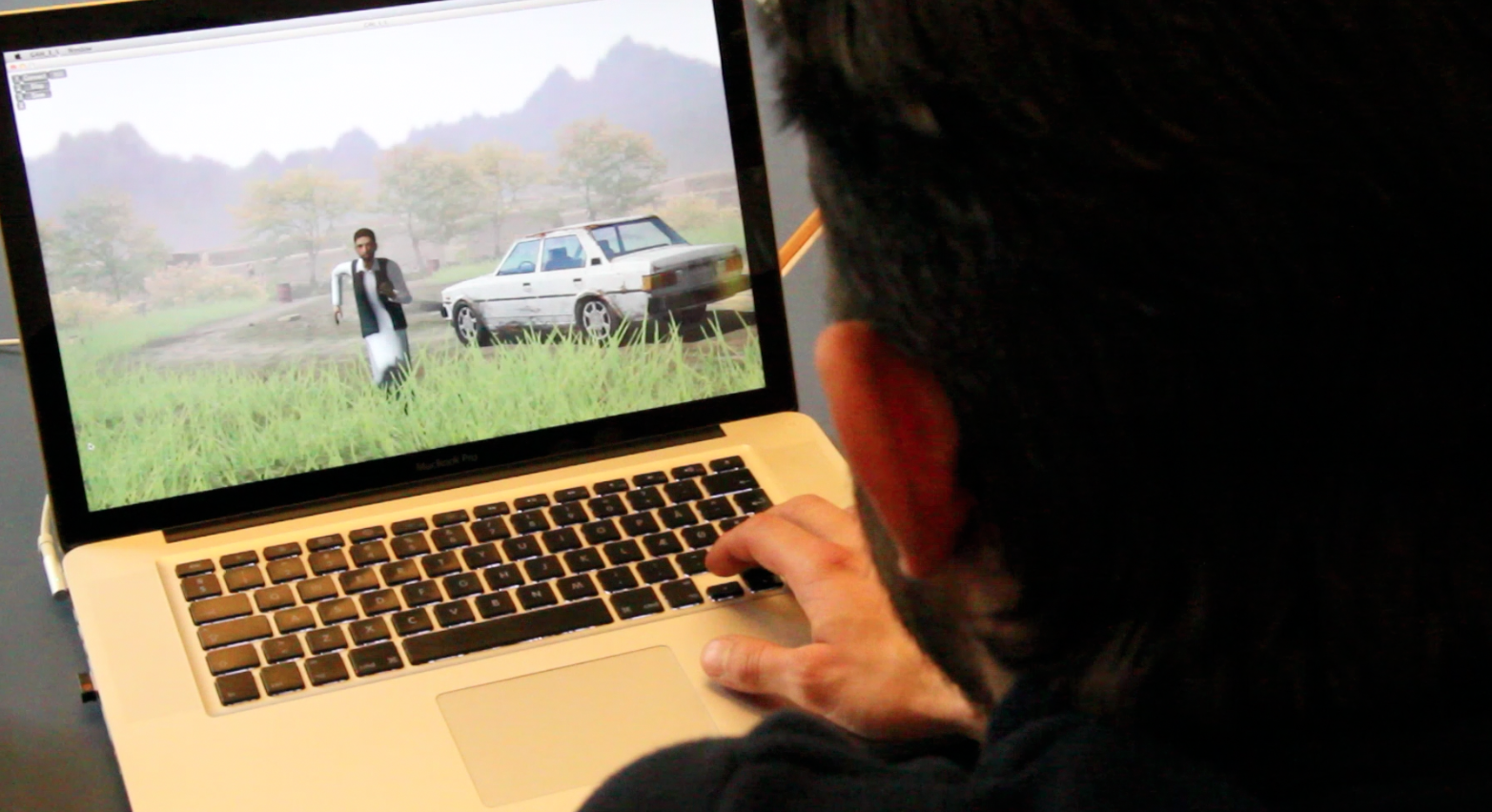}
    \includegraphics[width=1\linewidth]{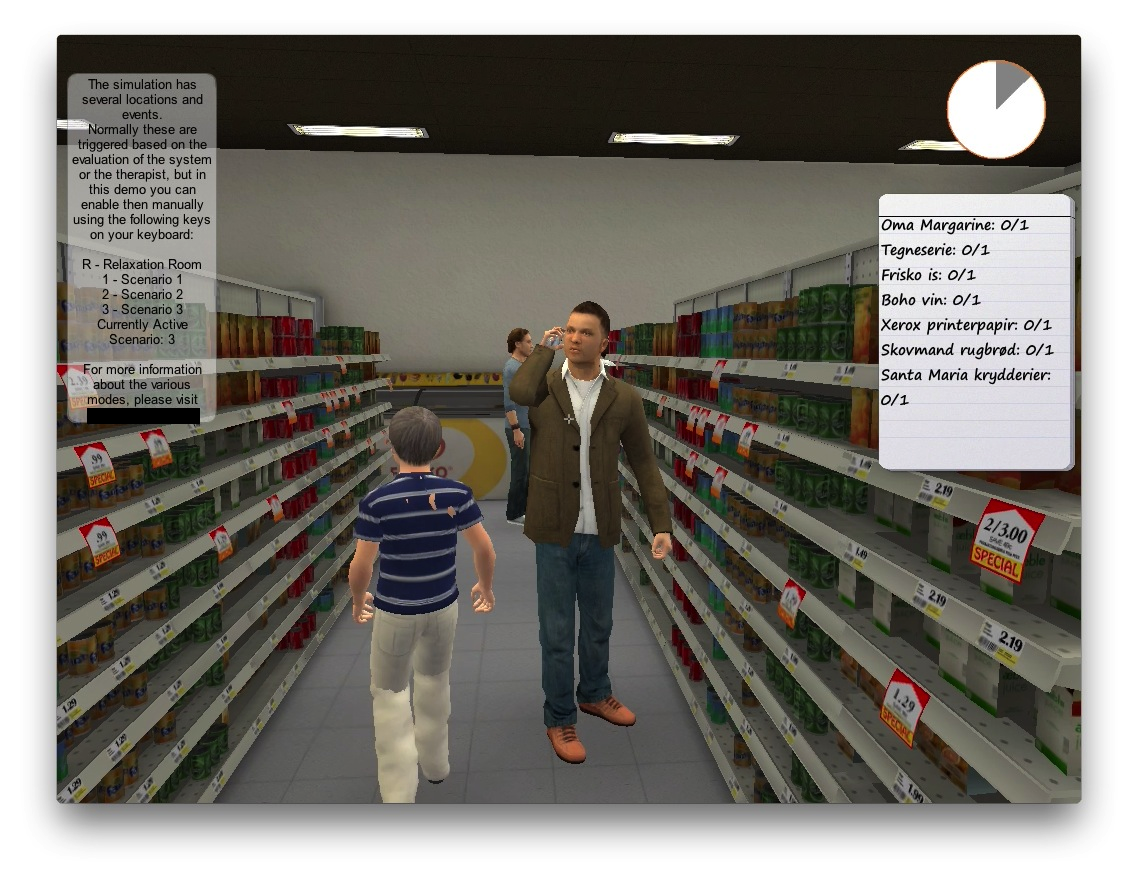}
    \caption{Screenshots of StartleMart \cite{holmgaard2015multimodal}; a biofeedback game designed as a virtual stress inoculation and exposure therapy tool.}
    \label{fig:startlemart}
\end{figure}

In this section, we will provide an indicative example of affect elicitors that have been used in a game that realises the game affection loop. In particular, we will outline the \emph{StartleMart} Post-Traumatic Stress Disorder (PTSD) game (see Fig. \ref{fig:startlemart}) which was designed and developed as a form of virtual exposure therapy \cite{holmgaard2015multimodal,holmgaard2015torank}. The game adapts to the level of stress of PTSD patients---as measured via their skin conductance---by triggering certain auditory and visual stimuli including war sounds, stressful social settings and war flashback moments. When it comes to the context of the game, StartleMart can be characterised as a training game with a health purpose (i.e. PTSD treatment) (\emph{genre}) that takes place in a supermarket, played in desktop (\emph{platform}) by a single player (\emph{number of players}). The game is partially observable (\emph{observability}) as it features a first-person view over a 3D level, it is deterministic (\emph{stochasticity}) and continuous (\emph{time granularity}), and the player actions are limited to moving around in a continuous environment and picking objects (\emph{action space}). The game context as described above acts as a mild stressor for patients suffering from PTSD. A pre-selected number of in-game stimuli---provided in the form of audio, visuals and video cut-scenes---act as personalised intense stress elicitors. More details about StartleMart can be found in \cite{holmgard2013stress,holmgaard2015multimodal,holmgaard2015torank}.

\section{Game Affect Sensing} \label{sec:sensing}

Moving on to phase 2 of the affective game loop in this section we will first survey the multiple ways we can sense how a player feels based on their manifested emotions (see Section \ref{sec:input}) and we will then survey the available methods we can obtain affect annotations and labels in games (see Section \ref{sec:output}). From an affect modelling perspective, we will first focus on the input of the model and then cover its output. 
Similarly to Section~\ref{sec:elicitation}, 
we start by introducing our taxonomy, then survey the corresponding literature and end with an indicative example that is presented in more detail.

\subsection{Sensing the Input} \label{sec:input}

Player emotional manifestations may be sensed through variations of gameplay patterns, alterations in a player's attention, level of focus, and changes in the player's physiology, facial expression, posture, and speech. Monitoring such alterations may assist in recognising and constructing the player's model. 

Any affect model of players relies on data of the manifested affective experience. Such data define the input of the affect model and can be obtained directly through the game engine or with the help of additional sensors. These sensors are either available through the various platforms (e.g. eye-tracking featured in a VR headset) or they are integrated into the game (e.g. a skin conductance sensor that interfaces with the game)---for a recent survey of physiological sensors in affective game research see also \cite{robinson2020let}. Following the player modelling taxonomies of \cite{yannakakis2018artificial,yannakakis2014emotion} we argue that sensing affect manifestations in games can be of the two main categories: \emph{gameplay} and \emph{objective} sensing. The two categories are detailed in the remainder of this section and summarised in Table \ref{tab:sensing}.
 
\subsubsection{Gameplay}

As games affect the player's cognitive processing patterns and cognitive focus our core assumption is that players' in-game behaviour is linked directly to their experience. As a result players' affect can be derived through the analysis of their in-game interaction patterns considering game context variables \cite{conati2002using,gratch2005evaluating}. Gameplay refers to anything a player does in a game environment which is collected via in-game logs of any type such as user interface selections, preferences, or in-game actions. In particular, gameplay includes any aspect that can be derived from the interaction between the player and the game directly. Such aspects---also broadly known as \emph{player metrics} \cite{el2013game,el2021game}---include detailed attributes of the player's behaviour which are based on interactions with game elements such as objects, non-player characters, and levels. Popular examples of attributes that are directly linked with gameplay include spatial locations of players and key events viewed as \emph{heat maps} \cite{drachen2013spatial} trajectories or aggregated descriptive data \cite{melhart2022again}, communication with other players or an audience \cite{melhart2020moment}, all the way to pixel colours of the game's footage \cite{makantasis2019pixels,makantasis2021pixels}.

A key limitation of gameplay is that the player affect is only observed \emph{indirectly}. For example, a player that shows limited interaction through their logged gameplay data could be either planning their next quest, talking to their friend over the phone or even feeling bored with the game. It is also important to note that affect models that are derived from gameplay data do not necessarily generalise across all players and games. Therefore, it is crucial that affect models are fine-tuned to the needs of players, and manifest gameplay experiences of player personas \cite{holmgaard2014personas} and ultimately individual players. One might even argue that gameplay data is not even relevant for particular games or players as the ad-hoc design of the gameplay attributes might not be useful for capturing certain aspects of player affect. 

\subsubsection{Objective} 

This category refers to any signals available as a response to in-game stimuli. In particular, objective ways of sensing affect include physiological signals---electrodermal activity (EDA), electrocardiogram (ECG), electromyogram (EMG), electroencephalogram (EEG) \cite{parsons2020classification,du2020emotion,vskola2021bcimanager}---camera-based signals including facial expression, head pose, gestures and body movement \cite{sugiyanto2020acquiface,greipl2021facial}, and verbal signals including speech and body movements. 

The analysis of physiological manifestations of psychology (i.e. psychophysiology) is well studied by now; see \cite{andreassi2000psychophysiology,calvo2009effect,yannakakis2016psychophysiology,ravajapsychophysiology,robinson2020let} among many. It is widely evidenced that arousing or tense events cause dynamic changes in both sympathetic (increase) and parasympathetic (decrease) nervous systems whereas low arousal (e.g. relaxing or resting) states increase the activity on the parasympathetic nervous system. Such activity may cause observable alterations, for instance in a player's facial expression, head pose, and EDA \cite{cacioppo2000psychophysiology,sharma2012objective}. A significant body of literature has focused on the relationship between a player's \emph{physiology} in response to their gameplay patterns \cite{tijs2008dynamic,nacke2008flow,mandryk2006using,mandryk2007fuzzy,rani2005maintaining,tognetti2010modeling,Drachen10SIGGRAPH,mcquiggan2007early}, relying on ECG \cite{yannakakis2010towards}, photoplethysmography \cite{yannakakis2010towards, tognetti2010modeling,zaib2022using}, EDA \cite{mandryk2006using, holmgaard2015rank,holmgard2013stress,holmgaard2015multimodal}, respiration \cite{tognetti2010modeling}, EEG \cite{nijholt2009bci,rebolledo2009assessing,alzoubi2009classification,parsons2020classification}, and eye movement \cite{asteriadis2009estimation,munoz2011towards}.  
In addition to physiology, the player's \emph{bodily expressions} can reveal real-time affective responses from the gameplay stimuli. Such input modalities, have been explored extensively in games and include facial expressions \cite{kapoor2007automatic,arroyo2009emotion,grafsgaard2011predicting,busso2004analysis,zeng2009survey}, muscle activation \cite{conati2009modeling,dennerlein2003frustrating}, body movement and posture \cite{asteriadis2009estimation,van2008exploring,kapoor2007automatic,d2009automatic,bianchi2002modeling}, haptics \cite{orozco2012role}, and gestures \cite{hoysniemi2005children}, 
On a higher level, objective input can be viewed largely as either verbal or non-verbal. Verbal input includes speech-based modalities \cite{vogt2005comparing,kannetis2009towards,juslin2005vocal,johnstone2000vocal,banse1996acoustic}, while no-verbal input may rely on text \cite{pang2008opinion,cook2012aesthetic,llano2014towards} or any of the above-mentioned modalities.

The limitations of objective inputs are several. First, most of the sensors are not available in a player's natural habitat (i.e. in the wild); second, most sensors are intrusive, thereby, affecting gameplay at large; third, the signals obtained are usually noisy due to environmental conditions and hardware limitations. We discuss these limitations in more detail in Section \ref{sec:discussion}.


\begin{table*}[!tb]
\caption{A Taxonomy of Game Affect Sensing}  
\centering
\begin{tabular}{|l||l|l|l|}
\hline 
\textbf{I/O} & \textbf{Category} & \textbf{Subcategory}  & \textbf{Sensing Options} \\ \hline \hline
 Input & Gameplay &  & Ad-hoc gameplay features, pixels, audio, preferences, in-game actions \\\cline{2-4}
  & Objective & Physiology & EDA, ECG, EEG, EMG \\
  &           & Camera-based & Facial Expression, Head Pose, Gestures, Body movement\\
  &           & Verbal & Speech\\ \hline\hline
Output & Subjective & & Human demonstrations of experience, annotations, questionnaires \\ \hline
\end{tabular}
\label{tab:sensing}
\end{table*}


\subsection{Sensing the Output} \label{sec:output}

The output of any affect model is usually a set of particular states (e.g. happy), a scalar (e.g. the emotional dimensions of arousal and valence), or an ordinal relationship (e.g. tension is higher now than before). Sensing the output is predominately achieved through a \emph{subjective} annotation process by the player themselves (first person) or by others (third person) such as game experience designers, peers or external observers. Those annotated labels ideally need to be as close to the ground truth of playing experience as possible \cite{yannakakis2018artificial}. Sensing the most reliable affect labels for players is a tedious and laborious task which defines a challenge in its own right. The area of affect annotation is long studied in the literature, yet there are still many open research questions left for the design of the ideal annotation collection protocol. First, who provides the labels (first or third person); second, is player experience represented as states or instead as intensity/magnitude; third, should annotations be provided in discrete time periods or continuously; fourth, should the annotators be asked to give a magnitude label (e.g. frustration is 0.8) or an ordinal relationship (e.g. frustration is higher in this level segment). Some of these questions are addressed in the remainder of the paper; particularly Sections \ref{sec:detection} and \ref{sec:dataCollection}.

\subsection{Indicative Examples}

\begin{figure}[!tb] 
    \centering
    \includegraphics[width=1\linewidth]{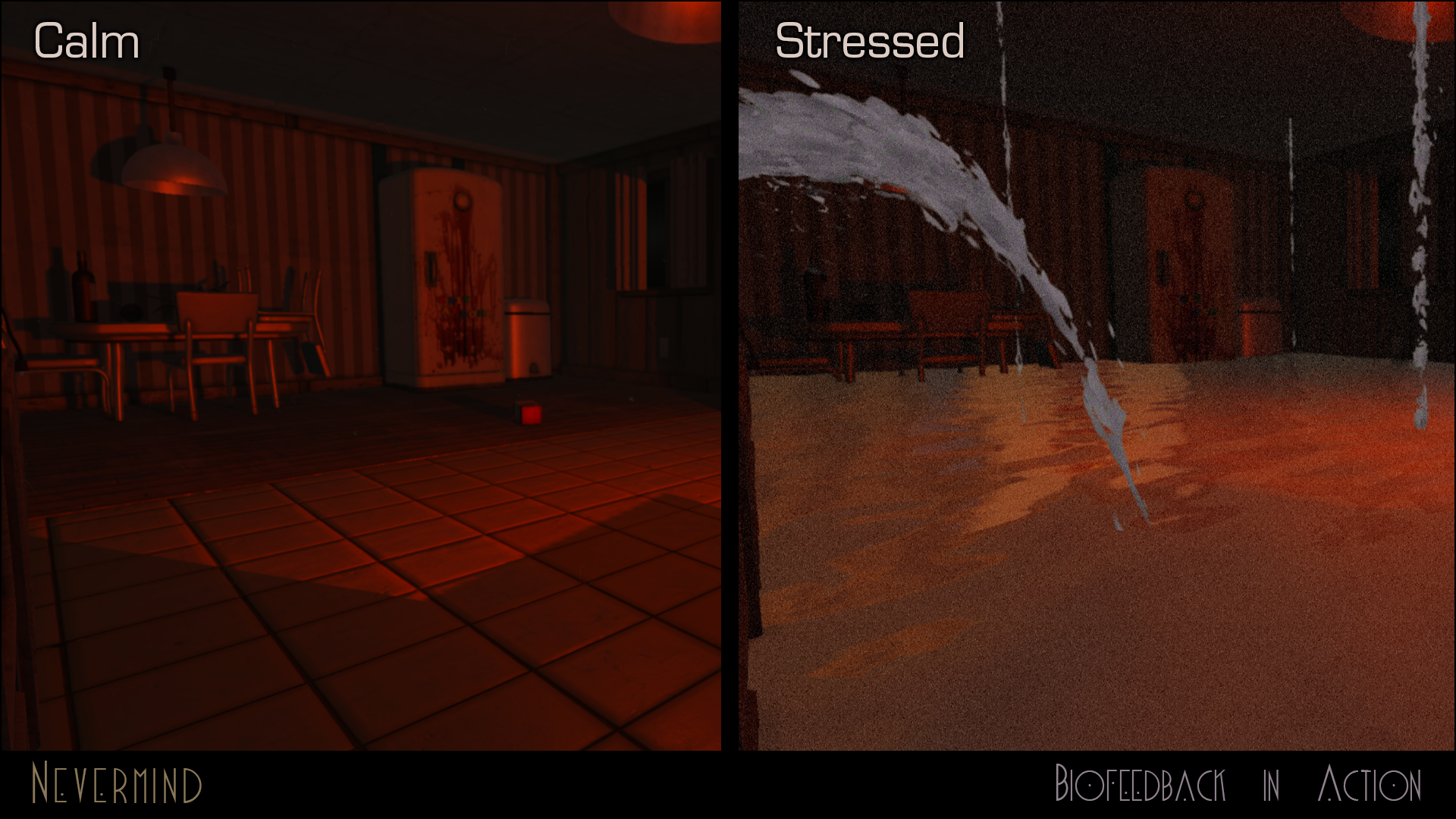}
    \caption{A screenshot of \emph{Nevermind} (Flying Mollusk, 2015) showcasing the elements of the game that can alter to elicit stress to the player. Image obtained from Erin Reynolds with permission.}
    \label{fig:nevermind}
\end{figure}

At the moment of writing, there are a number of examples of commercial games that utilise physiological input from players. One particularly interesting example is \emph{Nevermind} (Flying Mollusk, 2015), a biofeedback-based adventure horror game that adapts to the player's stress levels by altering the game's content including visuals and sounds (see Fig. \ref{fig:nevermind}) \cite{lobel2016designing}. A number of sensors which monitor the player's heart rate variability, skin conductance, facial expressions and gestures are available for affective interaction with the game. 

\begin{figure}[!tb] 
    \centering
    \includegraphics[width=1\linewidth]{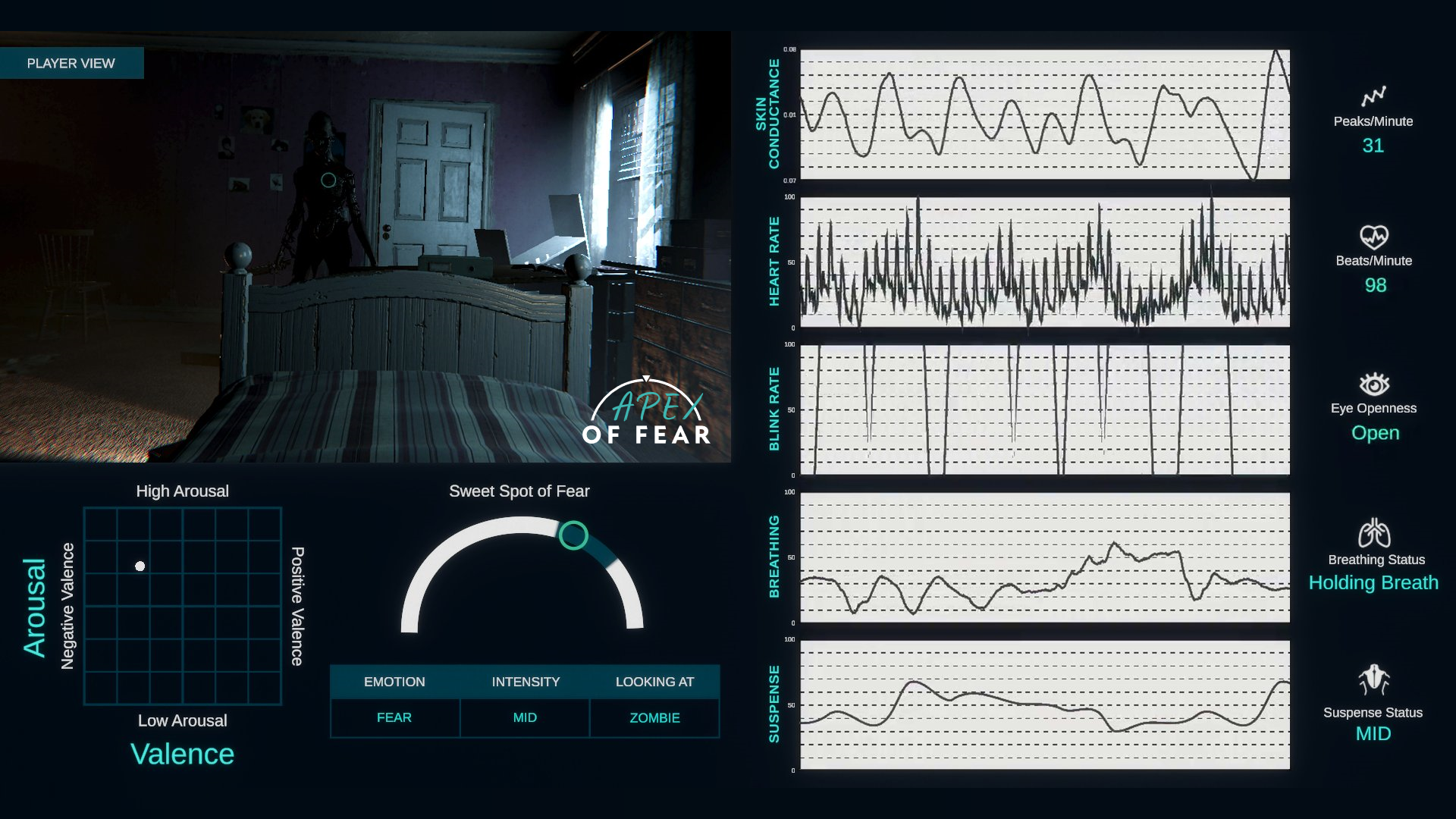}
    \caption{A screenshot of \emph{Apex of Fear} (2022). The dashboard showcases the player view and the multiple modalities available for sensing the players. Computational models of affect can be used to adapt the levels of fear according to a predetermined experience curve.}
    \label{fig:apex}
\end{figure}

Our second affect sensing example comes from a collaborative project between academic and industrial partners named \emph{Apex of Fear} (2022)\footnote{ \url{https://apexoffear.com/}}. \emph{Apex of Fear} is a VR horror game experience that adapts to its players' fear levels in real-time based on their psycho-physiological measures. The multimodal sensing capacities of \emph{Apex of Fear} include electrodermal activity, respiration, electrocardiography, electromyography, eye tracking, and in-game events (see Fig. \ref{fig:apex}). The game is currently in its beta testing phase through which user data is collected for constructing reliable models of in-game fear based on the physiological manifestations of players.

Both of these examples focus on \emph{sensing the input} rather than the \emph{output}. While surveys and continuous annotation tools to measure player experience are used extensively in games research \cite{melhart2021again} and game industry user research \cite{el2016game}, they rarely show up in commercial games. A possible explanation of why this happens is that most commercial games are presented as black boxes to the players. Game designers value immersive experiences highly \cite{tence2010challenge}, therefore, pointed surveys that measure the output of emotions directly can uncover the inner workings of game systems, revealing the smoke-and-mirror nature of games. Recent advancements, however, within 
large-scale language models (LLMs) point towards a promising direction incorporating human affect labels for shaping and defining large-scale affect models in games. As shown by Lambert et al. \cite{lambert2022illustrating}, reinforcement learning from human feedback can greatly enhance the performance and personalise the output of large foundation models such as GPT-3 \cite{floridi2020gpt} and GPT-4 \cite{openai2023gpt4}. Human feedback may take the form of like/dislike labels or preferences among options (i.e. generated texts and/or images).  

\begin{figure}[!tb]
    \centering
    \includegraphics[width=1\columnwidth]{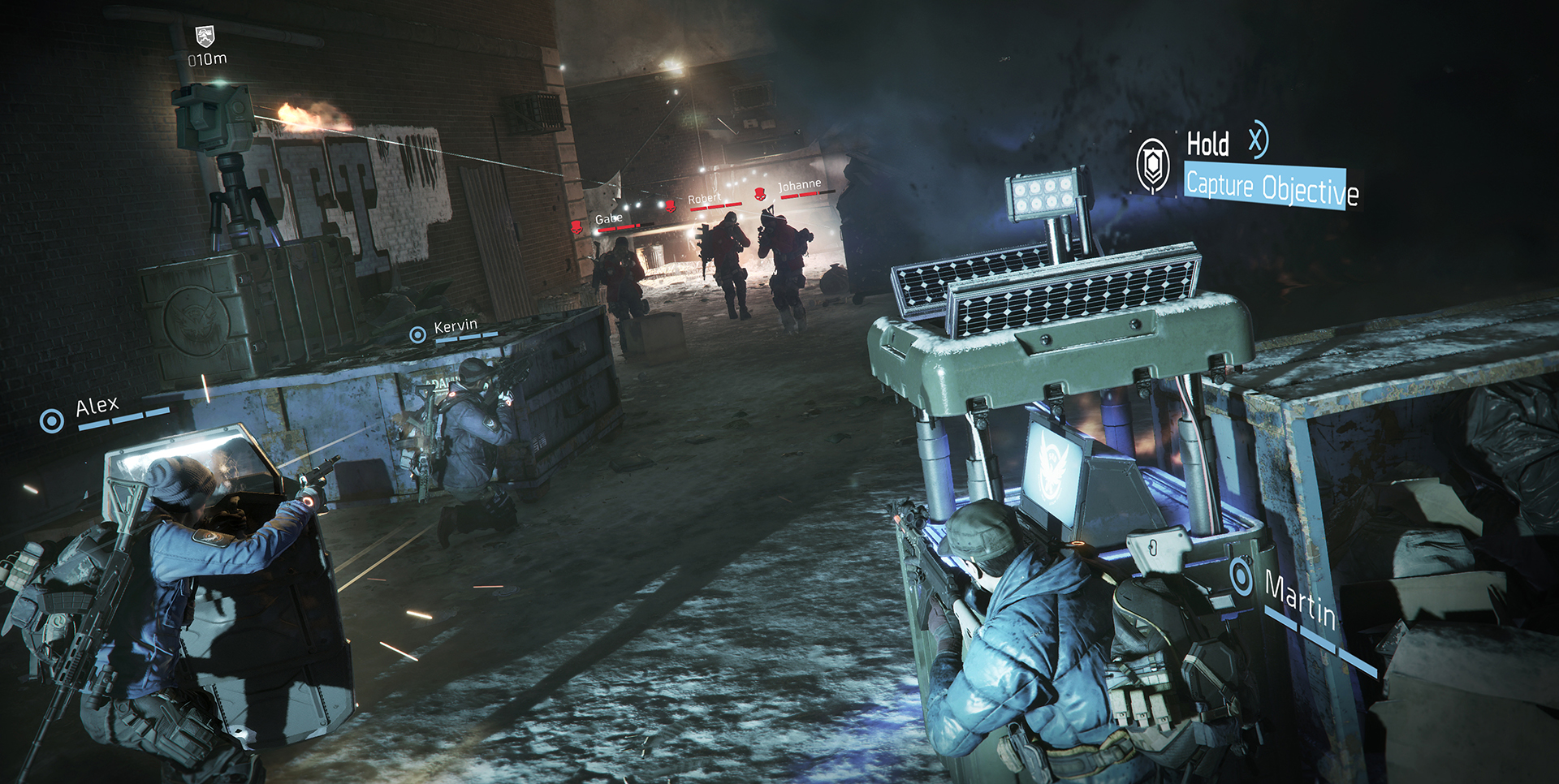}
    \caption{Game footage from \emph{Tom Clancy's The Division} (Ubisoft, 2016).}
    \label{fig:division}
\end{figure}

Even though general like/dislike labels already encode a form of affective feedback, detailed survey methods used in the games industry could inform future affect models in games. A good example of such a survey is the Ubisoft Player Experience Questionnaire (UPEQ) \cite{azadvar2018upeq} designed to assess the motivational drives of players. The UPEQ questionnaire has been used successfully to model player's motivation based on simple behavioral patterns of players in the role-playing third-person shooter game \emph{Tom Clancy's The Division} (Ubisoft, 2016) \cite{melhart2019your} (see Fig.~\ref{fig:division}).
 

\section{Game Affect Detection}\label{sec:detection}

\begin{table*}[!tb]
\caption{A Taxonomy of Game Affect Detection Methods}  
\centering
\begin{tabular}{|l||l|l|}
\hline
\textbf \textbf{Category} & \textbf{Subcategories} & \textbf{Learning Target}\\
\hline \hline
 Supervised Learning & Regression & Numerical values\\
 & Classification & Nominal categories\\
 & Preference Learning & Ordinal relations\\
 \hline
 Reinforcement Learning & Offline vs Online, Inverse Reinforcement Learning & Simulation of a human-like affective process \\ \hline
\end{tabular}
\label{tab:detection}
\end{table*}

In the third phase of the game affective loop, a computational model is requested to detect player affect based on the player modalities available. Because this affect model is trained to predict labels, the task of affect detection is largely viewed as a supervised learning paradigm \cite{calvo2009effect} through which measurable attributes of the player (model's input) are mapped to the player’s affect state (model's output). Any supervised learning method can be used for inferring such a mapping, including decision trees and random forests, support vector machines, and shallow or deep neural network architectures. 
The data type of the affect label available determines the output type of the model and, in turn, the machine learning approach that is applicable. Numerical data can be modelled using \emph{regression} methods whereas nominal variables, such as emotion categories or arbitrary bins of numerical data (e.g. high vs low values based on a split criterion) are modelled via \emph{classification} methods. Finally, ordinal observations (e.g. pairwise preferences or forced choices) can be trained via \emph{preference learning} methods. It is also possible to create ordinal observations from numerical values (e.g. change in score) or even from nominal values in some cases (e.g. arousal intensity of labelled emotions) \cite{yannakakis2018ordinal}. We detail the three supervised learning types below (see Table \ref{tab:detection}).

When the affect labels that need to be predicted are interval, affect modelling can be achieved via \emph{regression} algorithms including linear or polynomial regression, artificial neural networks and support vector machines. Modelling player affect via regression has certain limitations and will most likely yield unreliable models as regression assumes that the target value to be predicted follows a numerical scale. A number of comprehensive studies \cite{yannakakis2017ordinal,ovadia2004ratings,metallinou2013annotation,langley1985visual, yannakakis2015ratings,yannakakis2018ordinal,LNCS69740437} provide sufficient evidence against the use of regression for player experience modelling. Values obtained via magnitude-based annotation (e.g. ratings) should instead be converted to ordinal values via the \emph{second-order process} described in \cite{yannakakis2018ordinal}. 

If instead of numerical values player affect is represented as a set of classes (e.g. high vs low engagement) any \emph{classification} method is applicable for learning to predict affect. Classes can represent both aspects of player experience---e.g. \emph{excited} or \emph{frustrated} player---and aspects of playing behaviour such as quest completion times (e.g. low vs. high completion time). Classification is ideal for modelling player experience if discrete annotations of experience are available as target outputs \cite{LNCS69740014,LNCS69740155,giakoumis2011automatic,chen2019faceengage}. Converting numerical values to classes (e.g. convert arousal annotations between 0 and 1 to low vs. high arousal) might introduce data biases which, in turn, might prove to be detrimental for modelling player affect \cite{yannakakis2017ordinal,yannakakis2018ordinal,Martinez14Dont}.


Alternatively to regression and classification, \emph{preference learning} \cite{furnkranz2010preference,yannakakis2018ordinal} methods can learn to predict affect from ordinal data such as affect ranks or preferences. 
The target values in the preference learning paradigm provide information for the \emph{relative} relation between instances of the label we attempt to learn. For instance, labels are obtained by comparing two game levels in terms of engagement \cite{shaker2010towards} or they can be retrieved via ordinal processing of an arousal trace \cite{melhart2021again,melhart2019pagan}. Largely speaking, the data analysis of ordinal labels follows the first-order process described in \cite{yannakakis2018ordinal}. A large palette of algorithms is available for the task of preference learning including linear statistical models and non-linear approaches such as Gaussian processes\index{Gaussian process} \cite{chu2005preference}, deep and shallow artificial neural networks~\cite{martinez2013deep,burges2010ranknet,zacharatos2021emotion,makantasis2021affranknet+}, neuro-evolutionary methods \cite{pinitas2022rankneat}, and support vector machines \cite{joachims1998text}. Many of those methods are available in online accessible tools \cite{camilleri2019pyplt}.

Grounded in extensive studies available in the literature \cite{yannakakis2017ordinal,yannakakis2018ordinal} and supported by contemporary research \cite{makantasis2021affranknet+,pinitas2022rankneat} the selection of a supervised learning approach for modelling player affect becomes obvious. We argue that preference learning is a superior supervised learning method for player affect modelling, classification provides a good balance between simplicity and approximation of the ground truth of player experience whereas regression is based on numerical affect annotations which, in turn, yield models of questionable validity and reliability.

\subsection{Affect Detection as Reinforcement Learning}

The affect detection task is traditionally viewed through the lens of supervised learning. However, alternative approaches have emerged beyond this paradigm recently. In particular, methods from reinforcement learning have shown great potential for modelling player affect (see Table \ref{tab:detection}). The key motivation for the use of reinforcement learning (RL) for modelling player affect is that it can capture the relative valuation of affective states as encoded internally by humans during play \cite{tastan2011learning}. According to the RL perspective for modelling players, the derived RL policy can capture internal player states with no corresponding absolute target values such as decision-making, learnability, cognitive skills or style \cite{Aytemiz2021acting}. The player model that is built via an RL process is expected to offer a psychometrically-valid, abstract simulation of a human player's internal cognitive and/or affective processes. An agent equipped with such a model can be used to interpret human play, or featured in AI agents which can be used as playtesting bots or as believable human-like opponents \cite{holmgaard2014generative,barthet2021go,barthet2022generative}.

These non-traditional ways of detecting players' affect are still in their infancy with only a few studies existent in racing and Atari-like 2D games \cite{barthet2021go,holmgaard2014generative,holmgaard2014evolving}, racing games \cite{barthet2022generative}, serious games \cite{nguyen2014strategy} and first-person shooters \cite{tastan2011learning}.

\subsection{Indicative Examples}\label{sec:detection:examples}

We use racing games as the genre of the indicative examples we discuss in more detail here. Focusing on the game industry, the \emph{Drivatar} imitation learning system of the Forza Motorsport series (2005–2022, Microsoft Game Studios) is the longest-standing AI within a game title. Building on a set of simple rules and behavioural cloning techniques back in 2005 Drivatar has evolved to feature deep neural network approaches (in 2022) that imitate the way any player drives a car and simulate the player’s driving style\footnote{A video detailing the evolution of Drivatar is available here: \url{https://www.youtube.com/watch?v=JeYP9eyIl4E}}. Even though modelling in Drivatar relies on behavioural human demonstrations it is indicative of what player affect modelling can achieve in commercial-standard games if human affect demonstrations can be provided.  

\begin{figure}[!tb] 
    \centering
    \includegraphics[width=1\linewidth]{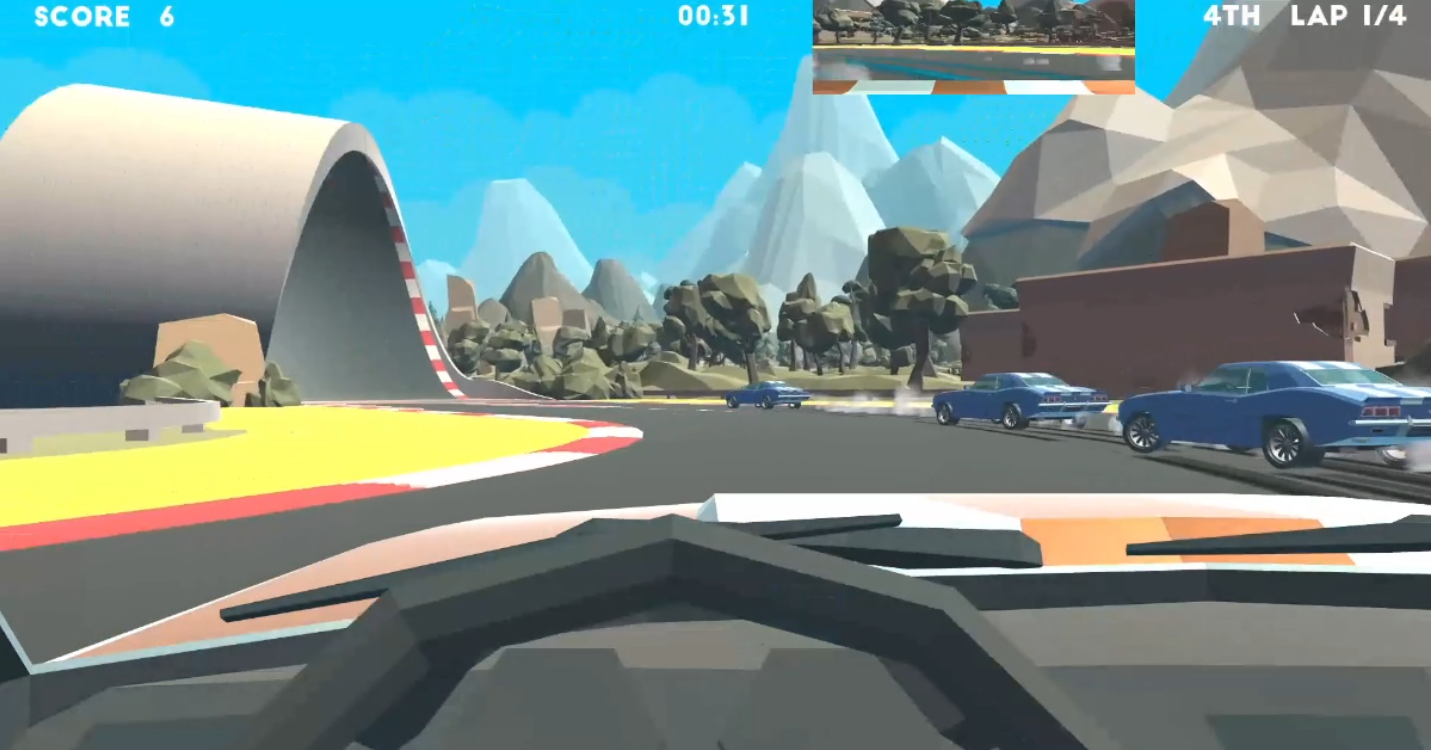}
    \includegraphics[width=1\linewidth]{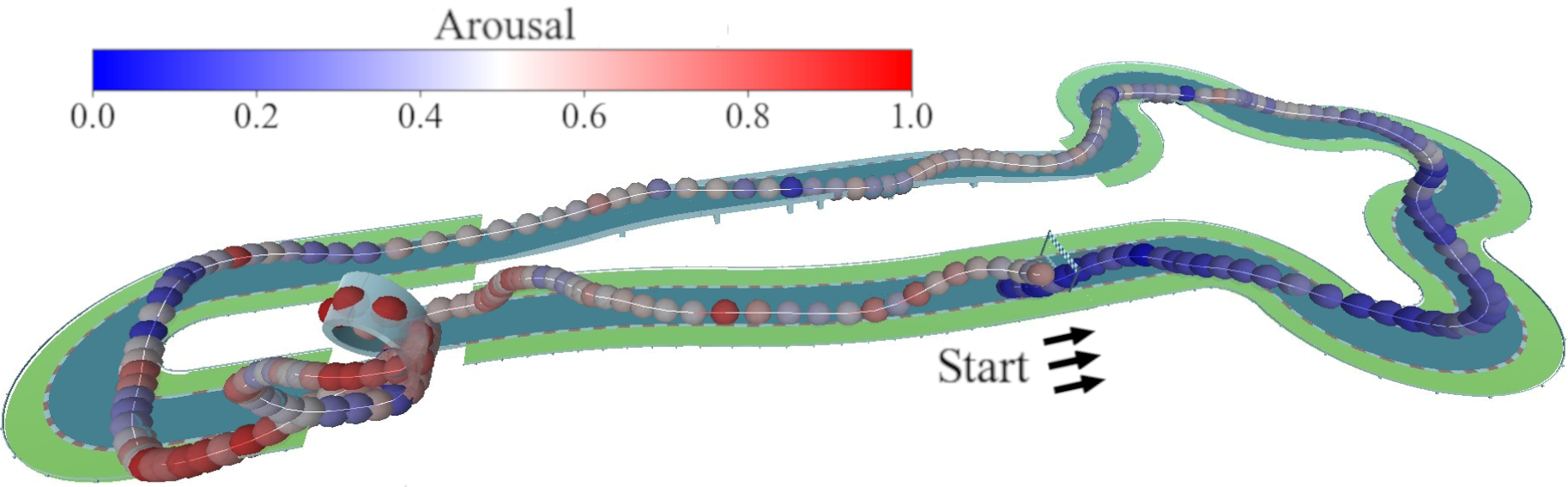}
    \caption{A screenshot of the Solid Rally racing game \cite{barthet2022play} (top) and the generated behavioural and arousal traces of an RL agent that learns how to play like and ``feel'' like a human expert player (bottom) \cite{barthet2021go,barthet2022play}.}
    \label{fig:goblend}
\end{figure}

Moving from the game industry to a recent research example in the area of affect modelling the work of Barthet et al.  \cite{barthet2021go,barthet2022generative,barthet2022play} is worth outlining. The authors in that series of papers import the highly successful reinforcement learning algorithm Go-Explore \cite{ecoffet2021first} to the field of affect modelling in their attempt to model both behavioural and affective patterns of play. The resulting algorithm, namely Go-Blend, is able to blend human demonstrations of behaviour and affect in a common representation thereby allowing for believable playtesting. The generative Go-Blend agents are able to play and ``feel'' like human players of racing games by imitating their play and arousal traces (see Fig. \ref{fig:goblend}). 

\section{Game Affect Adaptation} \label{sec:adaptation}

Computer games---as opposed to traditional media content such as images and videos---are interactive media that continuously react to the users' input. This interactivity can naturally accommodate mechanisms for real-time adaptation of game content aimed at adjusting player experience and realising affective interaction \cite{yannakakis2011experience}. One of the main reasons we can achieve meaningful affect-driven adaptation in games is because players are prepared for personalised experiences more than in any other form of human-computer interaction \cite{yannakakis2014emotion}. The relationship of players to adaptation mechanisms in games is highly dependent on their playing style, mood, experience, personality, and on the efficiency of the adaptation with regards to player needs.

The last phase of the game affective loop involves the adaptation of in-game elements for eliciting particular affective patterns. We refer to such elements as \emph{actionable} as they are linked and can directly affect player experience. Games may evolve and adapt to the player in many different ways and convey emotions through a variety of techniques and effects. Any adaptation process that will eventually close the affective interaction between the player and the game should be able to decide which stimulus (or playful experience) will be presented next, when it should be presented, which actionable game elements should be adjusted, and how \cite{yannakakis2014emotion,hutchings2019adaptive,hougaard2021willed}---for a recent survey on biofeedback interactions see \cite{navarro2021biofeedback}. Viewing affect-based adaptation from a high-level perspective it appears that the game can adjust its \emph{agents}---(or non-player characters) if those are available---or adjust its \emph{content} to the affective needs of its player(s). Both of these actionable in-game element types can be manipulated in ways that lead the player to become more emotionally involved with the game. We review these categories in the remainder of this section.

\subsection{Agents}

Several games feature agents or non-player characters that may act as opponents, collaborators, or assistants \cite{yannakakis2018artificial}. Independently of the type of agent and its role such agents might be required to express emotion during their interaction with the player(s) and elicit specific emotional patterns. Emotion expression can be achieved in a completely scripted manner (e.g. behaviour trees) all the way to machine-learned behavioural generation approaches (e.g. procedural personas). Approaches may rely upon popular emotion
agent architectures \cite{Paiva02,aylett2005fearnot,eladhari2008good}, underlying cognitive models \cite{gratch2004domain}, or personality trait models \cite{doce2010creating}.

It is important to note that emotion modelling plays a dual role when it comes to game agents: emotions both guide an agent's decision-making capacities but they also affect the expressions of different
emotional states (e.g. fear or sadness). Procedural animation is key for the latter with several impressive breakthroughs achieved for real-time animated characters \cite{perlin1985image} via generative systems \cite{habermann2021real}.

\subsection{Content}

Not all games feature non-player characters but all games have some form of content such as visuals, audio, game rules, game levels, and narrative; those content types can be defined as the creative facets available in games \cite{liapis2014computational}. An affect-driven adaptive process could in principle alter those creative facets independently or in an orchestrated manner \cite{liapis2018orchestrating,yannakakis2018artificial}. The area widely known as \emph{procedural content generation} (PCG) has offered several methods varying from simple search-based methods \cite{togelius2011search} all the way to deep learning-based approaches \cite{liu2021deep} for the task. Of particular importance for the scope of this paper is the experience-driven PCG
framework \cite{yannakakis2011experience,yannakakis2015experience}, which views game content as an indirect building block of
player affect and proposes mechanisms for synthesising personalised game experiences. Game content adaptation that may affect the emotional patterns of players can take the form of game rules \cite{togelius2008experiment}, difficulty \cite{or2021dl}, lighting \cite{xylakis2021architectural}, camera profiles \cite{yannakakis2010towards}, maps \cite{togelius2010towards}, levels \cite{shaker2013crowdsourcing}, tracks \cite{togelius06evolving}, narrative structures \cite{alvarez2022tropetwist}, and music \cite{hutchings2019adaptive}---among many other content types.


\subsection{Indicative Examples}

\begin{figure}[!tb] 
    \centering
    \includegraphics[width=1\linewidth]{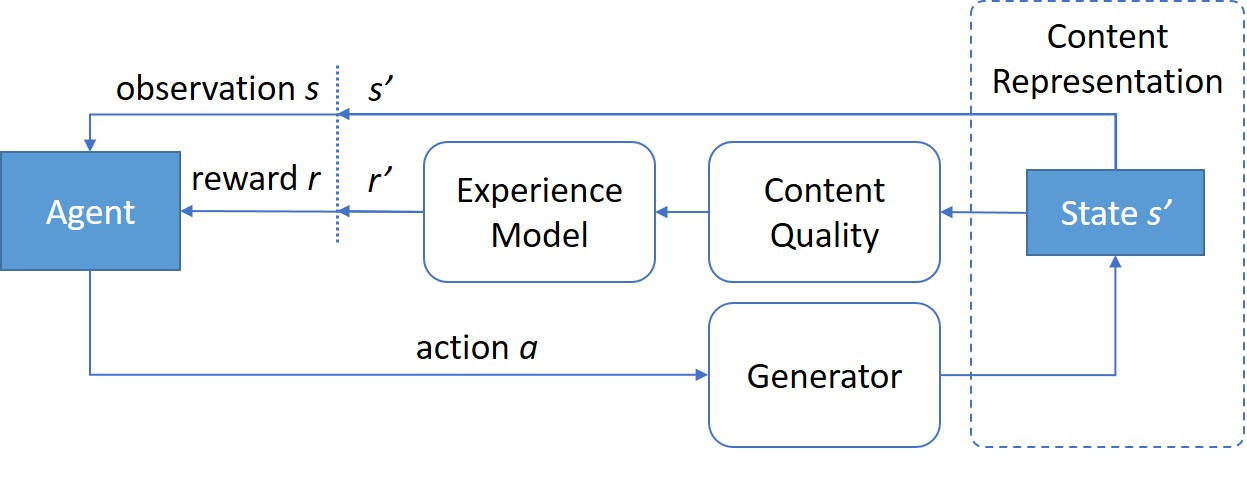}
    \includegraphics[width=1\linewidth]{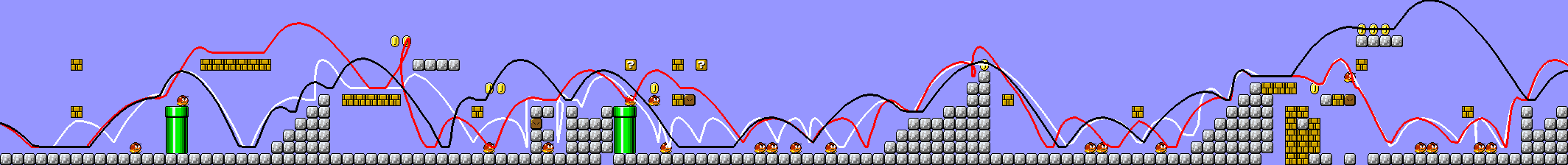}
    \caption{The ERDL framework (top) and an example of a \emph{Super Mario Bros} level it generates (bottom). The play traces of three dissimilar agents are overlaid on the level. EDRL moderates the divergence of game level and gameplay based on Koster's \emph{theory of fun} \cite{koster2013theory} and it designs levels endlessly via RL algorithms.
    }
    \label{fig:edrl}
\end{figure}

Affect adaptation in the games industry usually takes the form of Dynamic Difficulty Adjustment (DDA) \cite{yannakakis2018artificial,spronck2004difficulty,Andrade05}. In DDA, certain agent properties (e.g. enemy's health, speed and position) are predominately altered to match the skill of the player. Rubber-banding in racing games (e.g. in the \emph{Forza} series mentioned earlier) and difficulty scaling in games like the \emph{Resident Evil} series (Capcom, 1996-2023) are among the most popular methods for DDA in the games industry. While adaptation in games considers primarily a player's behaviour, there are games such as \emph{Left 4 Dead} (Valve, 2008) that follow the experience-driven PCG \cite{yannakakis2011experience} paradigm and, thereby, consider aspects of in-game tension and player's emotional intensity to adapt the game \cite{ambinder2011biofeedback}. The game's \emph{AI Director} observes the players' performance, follows a predetermined tension curve and modifies the in-game experience by altering a number of content types: the number and location of opponents (zombies), the pacing of the game and audiovisual effects\footnote{A video detailing the core aspects of the AI director is available here: \url{https://www.youtube.com/watch?v=Mnt5zxb8W0Y}}. The AI Director of \emph{Left 4 Dead} (Valve, 2008) defines a seminal use of adaptive AI in games and since its introduction in 2008 it has found multiple uses across games such as the \emph{Far Cry} series (Ubisoft, 2004-2021) and \emph{Watch Dogs 2} (Ubisoft, 2016).

As an indicative research example in affect-driven adaptation, we will cover Experience-Driven (ED) PCG via Reinforcement Learning (RL) or EDRL for short \cite{shu2021experience,wang2022fun}. EDRL is able to generate various facets of game content (e.g. game levels and gameplay patterns) that follow particular experience patterns via RL methods; see Fig. \ref{fig:edrl}. More specifically, the method was applied for the online and endless generation of \emph{game levels} and \emph{gameplay} patterns in the game of \emph{Super Mario Bros}. Inspired by Koster's \emph{theory of fun} \cite{koster2013theory}, reward functions were formulated as moderate degrees of level or gameplay divergence. The resulting \emph{multifaceted} EDRL is not only capable of generating \emph{fun} levels efficiently, but it is also robust with respect to dissimilar playing styles and initial game level conditions. 

\section{A Holistic view of Affective Game Computing}\label{sec:holistic}

This section provides a holistic overview of the field of affective game computing, focusing on state-of-the-art research, active areas, and research gaps in the field. Beyond the survey above, here we aim to show key examples of research across each phase of the \emph{affective loop}. While our survey gave an overview in light of the larger field of affective computing and user modelling, our holistic overview highlights papers with a videogame focus. First, we outline the methodology we used (Section \ref{sec:method}) and then we analyse our findings (Section \ref{sec:observations}).

\subsection{Methodology}\label{sec:method}

Sections~\ref{sec:elicitation}---\ref{sec:adaptation} pull knowledge from the last two decades of research in affective game computing, This section, instead, complements our literature review with a survey of recent papers (i.e. past 4 years) from the primary publication venues of affective game computing research: the 
\textit{IEEE Transactions on Games} (ToG) and 
\textit{IEEE Transactions on Affective Computing} (TAC) journals, 
and the \textit{IEEE Conference on Games} (CoG), the 
\textit{Conference on the Foundations of Digital Games} (FDG), the
\textit{Computer-Human Interaction in Play Conference} (CHI-Play), 
and the \textit{Conference on Affective Computing and Intelligent Interaction} (ACII).

Table~\ref{tab:summary} shows the outcome of this survey complemented with seminal works over the last 20 years of affective game computing. The table is organised based on the core aspects of the affective game loop---\textit{elicitation, sensing, detection}, and \textit{adaptation}. While many of the presented papers utilise multiple---if not all---aspects of the affective loop, we have selected papers that focus on a specific phase of the loop with the corresponding high-level categories defined in earlier sections of the paper. 

\begin{table*}[tb!]
    \centering
    \caption{Summary of surveyed studies associated with the four core phases of the affective game loop---Elicitation, Sensing, Detection, and Adaptation. The studies are placed under each of the four phases based on their primary focus within the affective loop. They are also placed under one of the main categories of each phase based on their research emphasis.}
    \begin{tabular}{|p{1.5cm}||p{3cm}|p{9cm}|}
        \hline
        \textbf{Phase} & \textbf{Category} & \textbf{Papers}\\
        \hline
        \hline
        Elicitation & Context & \cite{graja2020impact} \cite{melhart2022again} \cite{pallavicini2018effectiveness} \cite{born2021exergaming} \cite{pisalski2020influencing} \cite{somarathna2022virtual} \cite{pallavicini2019gaming} \cite{canossa2020hold} \cite{ravajapsychophysiology} \\
        \cline{2-3}
         & AI Agent & \cite{ravaja2016virtual} \\
         \cline{2-3}
         & Content & \cite{reetz2021nature} \cite{colombo2021psychometric}	\cite{lara2018induction} \cite{ogawa2020auditory}	\cite{lopes2017modelling} \cite{liapis2018orchestrating}	\cite{pedersen2010modeling} \cite{holmgaard2015multimodal}\\
        \hline \hline
        Sensing & Gameplay & \cite{yannakakis2014emotion} \cite{el2021game} \\
        \cline{2-3}
         & Physiology & \cite{robinson2020let} \cite{vskola2021bcimanager} \cite{du2020emotion} \cite{zafar2018gaming} \cite{Drachen10SIGGRAPH} \cite{navarro2021biofeedback} \\
         \cline{2-3}
         & Camera-based & \cite{sugiyanto2020acquiface} \cite{munoz2011towards} \cite{grafsgaard2011predicting} \\
         \cline{2-3}
         & Verbal \& Speech & \cite{kannetis2009towards} \cite{abramov2021analysis} \\
         \cline{2-3}
         & Subjective measures & \cite{melhart2020moment} \cite{azadvar2018upeq} \cite{yannakakis2015grounding} \\
        \hline \hline
        Detection & Classification & \cite{parsons2020classification} \cite{makantasis2021affranknet+} \cite{giakoumis2011automatic} \cite{chen2019faceengage} \\
        \cline{2-3}
         & Preference Learning & \cite{yannakakis2015ratings} \cite{melhart2021towards} \cite{Martinez14Dont} \cite{shaker2010towards} \\
         \cline{2-3}
         & Reinforcement Learning & \cite{holmgaard2014personas} \cite{barthet2021go} \cite{barthet2022generative} \cite{Aytemiz2021acting} \cite{barthet2022play} \\
        \hline \hline
        Adaptation & Content & \cite{hougaard2021willed} \cite{lobel2016designing} \cite{or2021dl} \cite{zaib2022using} \cite{hutchings2019adaptive} \cite{parnandi2017visual} \cite{liu2021deep} \cite{alvarez2022tropetwist} \cite{xylakis2021architectural} \cite{shaker2013crowdsourcing} \cite{togelius2011search} \cite{shu2021experience} \cite{wang2022fun}\\
        \cline{2-3}
        & Agents & \cite{habermann2021real} \cite{de2022adaptive}\\
        \hline
    \end{tabular}
    \label{tab:summary}
\end{table*}

\subsection{Analysis of Findings}\label{sec:observations}

In this section we discuss overall observations regarding the papers identified under each one of the four main phases of the affective loop (Table \ref{tab:summary}.) \textit{Elicitation} encompasses papers that focus on the emotional impact of games and their measurement. We divided these papers based on the game facet they consider. While many studies seem to put an emphasis on the game context, most of them focus on specific genres and platforms. In particular, the horror genre \cite{lopes2017modelling,graja2020impact,pallavicini2018effectiveness} provides a popular research test-bed, possibly due to the visceral emotions that emerge during consuming horror media. VR and AR games are also popular for studying the impact of gameplay context on player experience  \cite{born2021exergaming,pisalski2020influencing,somarathna2022virtual}. Even though there are some studies dedicated to the ways the action space of the game affects the player experience \cite{canossa2020hold,ravajapsychophysiology}, there are no studies found that focus on the impact of the number of players, observability, stochasticity, and time granularity on player affect. When it comes to AI agents, there is surprisingly little research dedicated on how these agents can be used for emotion elicitation despite the field of believable non-player characters \cite{pacheco2018studying,even2017analysis} being very active. Unsurprisingly, most studies with a primary focus on affect elicitation are investigating the impact of the game content and the game environment \cite{reetz2021nature}, virtual objects \cite{colombo2021psychometric}, sounds \cite{ogawa2020auditory}, game events \cite{ravajapsychophysiology} or multiple facets of content \cite{liapis2018orchestrating} on affect elicitation. Summing it up, it appears that there are two major research gaps identified in game affect elicitation. First, there is lack of fundamental research into how components of the game context affect the player experience. Second, there is an untapped opportunity to utilise the research that has gone into creating more emotive and human-like agents for emotion elicitation in affective game computing studies.

\textit{Sensing} appears to be a well-researched phase, with a special focus on physiology and peripheral signals (e.g. \cite{robinson2020let,navarro2021biofeedback}). Similarly, thanks to the ubiquity of web cameras, camera-based methods are also popular \cite{munoz2011towards,sugiyanto2020acquiface}. While gameplay on its own can be a strong predictor of the player experience---and it is often the only modality available in the wild \cite{el2021game}---most studies in affective game computing aim for a multimodal approach instead. Finally, the field of voice-based sentiment and emotion analysis is overshadowed by more traditional user modalities such as physiological signals. This latter finding is surprising as in the world of eSports \cite{abramov2021analysis} and streaming services, access to player voice and commentary are easier than ever. Regarding subjective measures, game affect sensing appears to adapt traditional affective computing labelling techniques to games---from annotation tools \cite{melhart2019pagan,melhart2021again} to Likert-like surveys \cite{azadvar2018upeq}---but there is relatively little research effort put on the design of game-specific annotation tools---with VR and AR platforms being the exception \cite{xue2021rceavr} due to the involvement required from players.

Almost all studies presented involve some form of \textit{detection} as it is the machine learning aspect that often aids the evaluation of other components of the affective loop. Most studies with a primary focus on detection traditionally involve supervised learning. While regression is still used from time to time for predicting aspects of player behaviour (e.g. purchase decisions \cite{sifa2015predicting}), when it comes to affect modelling, the vast majority of the surveyed research studies have a preference for classification methods. This is despite a long line of research advocating for the ordinal processing and modelling of emotions \cite{yannakakis2018ordinal,yannakakis2017ordinal}. More interestingly, we observe a contemporary increase of interest in affect-driven reinforcement learning \cite{barthet2021go, barthet2022generative, Aytemiz2021acting}. Seeing how imitation learning algorithms are already being used in the games industry (see Section~\ref{sec:detection:examples} about \textit{Forza Motorsport}), we expect a rapid growth of interest in RL-based affect detection.

Finally, we take a look at studies mostly linked to the phase of \textit{adaptation} (bottom row of Table~\ref{tab:summary}). One could argue that most commercial games that use some type of player-based adaptation employ a form of the affective loop. Similarly in academia, the research area of game affect adaptation is getting some serious traction in recent years. As Table~\ref{tab:summary} shows, there are plenty of research projects that focus on affect adaptation and intelligent interaction based on sensing and emotion detection. 
Unsurprisingly, similarly to studies focusing on elicitation, most studies involving game affect adaptation focus on the game content instead of interactive agents \cite{habermann2021real}. While the holistic approach of experience-driven procedural content generation allows for more in-game adaptation opportunities \cite{yannakakis2011experience}, the lack of emotive agents in affect adaptation reveals a notable gap in the research field. Despite the successful affect-adaptive methods showcased in the literature, the widespread adoption of these techniques in commercial-standard games is still in its infancy. One of the biggest obstacles in this regard appears to be the field's reliance on often intrusive biosensors. While most physiological sensors can provide valuable multimodal data, such data are often very hard to reliably capture in the wild. We expect that future research avenues will investigate methods that would allow rapid deployment in games via camera-based technology or via the use of user-agnostic models---e.g. learning from in-game footage pixels or via privileged information \cite{makantasis2021pixels,makantasis2021privileged}.

This section completes the survey and taxonomy of affective game computing. In the second part of this paper (as initiated in Section \ref{sec:dataCollection}) we survey the tools available for reliable data collection in games and review the various game affect corpora that are currently available (Section \ref{sec:corpora}).

\section{Collecting Affective Data in Games}\label{sec:dataCollection}

Many tools for affective computing research can support games user research applications; however, games impose a number of unique challenges. Indicatively, different gaming interfaces provide different affordances both for play and data collection. Therefore, any dissimilarities across game controls and platform configurations would likely yield discrepancies in the collected telemetry and peripheral signals. 
Keeping a focus on affective data collection the remainder of this section provides an overview of popular gaming interfaces (Section \ref{sec:gaming}), and available tools for sensor data collection (Section \ref{sec:devices}) and annotation (Section \ref{sec:annotationTools}).

\subsection{Gaming Interfaces and Telemetry} \label{sec:gaming}

When it comes to selecting a game interface for an affective game computing experiment there are several aspects that need to be considered. \emph{Desktop} computers provide the most straightforward way of data collection. Since players have to be seated in front of a computer---operating with a keyboard and a mouse---they are generally less obstructed. This makes desktop setups ideal for laboratory studies. Designing and running experiments from a desktop computer also implies an easier integration between the experimental games and the research software. Due to the generally similar setup, crowd-sourcing methods for data collection can be relatively reliable. On the flip side, desktop computers come in many different hardware configurations, which can affect the game performance, user experience, and quality of the data. Some of the available \emph{consoles}, instead, such as the \emph{Microsoft XBox}\footnote{\url{https://www.xbox.com/}}, the \emph{Sony PlayStation}\footnote{\url{https://www.playstation.com/}}, and the \emph{Nintendo Switch}\footnote{\url{https://www.nintendo.com/switch/}} include specialised gaming hardware. Due to the standardised hardware and software, the collection of data related to player experience on these platforms can be more consistent. These specialised systems, however, also pose a major limitation as it is generally hard to integrate console software with research-based hardware such as biosensors; the latter has often to be operated from a separate desktop computer. 

As both desktop and console games are generally played by sitting in front of a monitor, the collection of traditional in-game behavioural telemetry, facial features, eye tracking information, and biosignals is relatively easier. In contrast, games that require full body movement are harder to integrate with more intrusive biosensors---especially the ones that involve electrodes and wires. Nevertheless, these games (and the platforms they are played on) provide unique affordances. Thanks to advances in computer vision and wearable technology, several exer-, VR-, and mobile games already integrate peripheral sensors into their systems.

\emph{Mobile} games use regular smartphones as their platform. Due to the interactions afforded by these phones, the collected telemetry can cover an input space different to traditional consoles and desktop computers---such as tapping patterns and gestures. Many smart accessories are readily interfacing with smartphones, making it easy to collect peripheral data from different biosensors---such as photoplethysmography obtained via a smartwatch. The drawback of the platform is the uncertainty and obfuscation of the gameplay context. Notably, mobile games usually are played across many different environments for very short periods of time or with irregular breaks during the experience. 

Exer(cise)-games exist on both console and mobile platforms. They usually rely on peripheral sensors that collect body movement and biosignals \cite{born2021exergaming}. The particularity of such games with regards to affective game computing is that they often employ a number of sensors to capture various modalities of the player. As players of exergames tend to (and have to) move their body throughout the game, the placement and arrangement of any physiological sensor poses a major challenge. Beyond the obstructed play, the motion and experimental artefacts that are embedded in the collected data make data cleaning and processing a non-trivial task \cite{yannakakis2008entertainment}.

Beyond console, desktop and mobile platforms, \emph{VR and VR} platforms---which become increasingly popular over the last few years---open a promising avenue for rich multimodal data collection in games; the \emph{Apex of Fear} (2022) project covered earlier is one such example. VR Headsets such as the \emph{Meta Quest}\footnote{\url{https://www.meta.com/quest/products/quest-2/}}, \emph{Valve Index}\footnote{\url{https://store.steampowered.com/valveindex}}, \emph{HTC Vive}\footnote{\url{https://www.vive.com/}}, and HP Omnicept headset\footnote{\url{https://www.hp.com/us-en/vr/reverb-g2-vr-headset-omnicept-edition.html}} may provide a unique 360$^{\circ}$ immersive experience \cite{somarathna2022virtual}. The VR headsets available generally block almost all obstructions  coming from the physical environment and many can be used for multimodal data collection including gaze and physiology (see more discussion in Section \ref{sec:discussion}). 

Generally speaking, affect data collection based on passive affect elicitors, such as images, is usually combined with signals obtained via physiological and other peripheral sensors. In games, however, it is a dominant practice to collect data from in-game \emph{telemetry} in addition to any other sensor data and independently of the game platform used. This type of behavioural data can be very powerful, especially in multimodal applications. The form of such data, however, highly depends on the given game and purpose of collection. While some game engines offer standardised methods to collect various types of user data, these methods often focus on data relating to monetisation (see for example the Unity Analytics\footnote{\url{https://unity.com/products/unity-analytics}}). 



\subsection{Sensing Devices and Tools} \label{sec:devices}

One of the most comprehensive surveys of affective computing research and development tools to date shows that even though there is a continuous development of new tools, most studies end up using custom-made solutions \cite{hussain2014tools}. This is also largely true for data collection, labelling, data processing, and machine learning platforms and tools, and it makes intuitive sense, especially when it comes to collecting data from games.

Collecting physiological sensor data is typically achieved through a vendor-based solution. The market is filled with reliable solutions for EEG, EDA, HR, and eye-tracking sensors. Some of the popular vendors include Emotiv\footnote{\url{https://www.emotiv.com/}}, Empatica\footnote{\url{https://www.empatica.com/}}, Intel Realsense\footnote{\url{https://www.intelrealsense.com/}}, Plux\footnote{\url{https://www.pluxbiosignals.com/}}, Polar\footnote{\url{https://www.polar.com/}}, Shimmer\footnote{\url{https://shimmersensing.com/}}, and Tobii\footnote{\url{https://www.tobii.com/}}, among many others. Most of these vendors offer processing software as well, although open-source solutions do exist. A recent survey on the field of affective game research showed that the most popular modalities employed include heart activity, followed by facial recognition and EDA \cite{robinson2020let}. Beyond specialised sensors, conventional cameras are also often used to collect recordings of subjects. Vendor-based solutions exist to process such multimodal data (i.e. Affedex \cite{mcduff2016affdex} for facial recognition); researchers, however, may prefer to use open source tools such as OpenCV\footnote{\url{https://opencv.org/}} for general computer vision tasks or OpenFace \cite{baltrusaitis2018openface} for facial recognition. 

While most affective computing methods can be transferred to game research without any major hiccups, sensing technology, in particular, brings two core challenges to the table. First, playing is usually physically active---even if the user plays on traditional platforms using a controller---which means that any sensor used for data collection has to afford a degree of comfort and free movement. Second, because games generally impose a high level of cognitive load on the user, there is an expected loss of affect expressivity. Due to this limitation, some more popular detection methods---such as face-based affect recognition---might be less relevant when applied to games \cite{roohi2018neural,shaker2013fusing}.

\subsection{Annotation Tools}\label{sec:annotationTools}

\begin{table*}[!tb]
    \centering
    \caption{A Review of Annotation Tools} 
    \begin{tabular}{|L{2cm}|L{3cm}|L{3cm}|L{4cm}|}
    \hline
    \textbf{Tool} & \textbf{Dimensions} & \textbf{Label} & \textbf{Installer} \\
    \hline \hline
    FeelTrace \cite{cowie2000feeltrace} & 2 dimensions (arousal-valence) & bounded continuous circumplex & N/A \\
    \hline
    ANVIL  \cite{kipp2001anvil} & Categorical labels & discrete labels & Standalone installer \\
    \hline
    ELAN  \cite{wittenburg2006elan} & Categorical labels & discrete labels & Standalone installer \\
    \hline
    EmuJoy  \cite{nagel2007emujoy} & 2 dimensions (arousal-valence) & bounded continuous & Standalone installer \\
    \hline
    AffectButton \cite{broekens2013affectbutton} & 3 dimensions (pleasure-arousal-dominance) & bounded continuous & Standalone or online \\
    \hline
    ANNEMO \cite{ringeval2013recola} & 1 dimension (configurable) & bounded continuous & Node.js package \\
    \hline
    GTrace \cite{cowie2013gtrace} & 1 dimension (negative to positive) & bounded continuous & N/A \\
    \hline
    CARMA \cite{girard2014carma} & 1 dimension (negative to positive) & bounded discrete (configurable) & Installer (requires MATLAB\footnote{\url{https://www.mathworks.com/products/matlab.html}}) \\
    \hline
    AffectRank \cite{yannakakis2015grounding} & 2 dimensions (arousal-valence) & discrete circumplex & Adaptation through PHP and JavaScript \\
    \hline
    RankTrace \cite{lopes2017ranktrace} & 1 dimension (tension) & unbounded continuous & C\# source (pre-built version available; requires VLC\footnote{\url{https://www.videolan.org/}}) \\
    \hline
    DARMA \cite{girard2018darma} & 2 dimensions (configurable) & bounded continuous (optional circumplex) & Installer (requires MATLAB, VLC, and a joystick) \\
    \hline
    NOVA \cite{heimerl2019nova} & 1 dimension or categorical  (configurable) & bounded continuous or discrete labels & Standalone installer \\
    \hline
    PAGAN \cite{melhart2019pagan} & 1 dimension (configurable) & unbounded and bounded continuous and discrete binary (configurable) & No installation (online) \\
    \hline
    RCEA \cite{zhang2020rcea} & 2 dimensions (arousal-valence) & bounded continuous circumplex & N/A \\
    \hline
    RCEA-360VR \cite{xue2021rceavr} & 2 dimensions (arousal-valence) & bounded continuous & Python package \\
    \hline
    \end{tabular}
    \label{tab:annotation}
\end{table*}

In this section, we review the available annotation tools that could be used for collecting affect labels in games. We also discuss the appropriateness of these tools for affective game computing. Table \ref{tab:annotation} summarises our survey on popular annotation tools that were introduced during the last two decades\footnote{N/A indicates that an installer is ``not-available''.}. One of the major trends of the affective computing field is the shift from discrete and more complex annotation methods towards simpler, continuous labelling techniques. Early annotation tools such as ANVIL \cite{kipp2001anvil} or ELAN \cite{wittenburg2006elan} (released in 2001 and 2006, respectively) focused on measuring discrete categorical emotions. While the Natural Language Processing field still uses such tools for labelling speech and text and certain annotation software still offers categorical labelling options (i.e. NOVA \cite{heimerl2019nova}) the field of affective computing started to shift away from such tools relatively early. With the advent of machine learning, moment-to-moment affect modelling became feasible---and with it, continuous annotation tools started to be used more widely. Traditional annotation tools such as FeelTrace \cite{cowie2000feeltrace} (released in 2000), AffectButton \cite{broekens2013affectbutton} (first released in 2009), and AffectRank \cite{yannakakis2015grounding} (released in 2015) measure multiple dimensions of affect at once---in some cases up to three dimensions. Annotation tools of that period were inspired by the \emph{Circumplex Model of Emotions} \cite{russell1980circumplex} and generally used to label the arousal-valence-dominance spectrum. 

While the two-dimensional annotation scheme is still popular nowadays we note a shift towards one-dimensional labelling tools (from 2013 onwards) including GTrace \cite{cowie2013gtrace} (released in 2013), ANNEMO \cite{ringeval2013recola} (released in 2013), CARMA \cite{girard2014carma} (released in 2014), RankTrace \cite{lopes2017ranktrace} (released in 2017), and PAGAN \cite{melhart2019pagan} (released in 2019). Similarly to FeelTrace in multi-dimensional labelling, GTrace became the new foundation for one-dimensional annotation tools. The bounded, continuous annotation method quickly spread and became popular in human-computer interaction and affective computing studies \cite{baveye2015deep,muller2015emotion,dellandrea2016mediaeval,dhamija2018automated}. The first new annotation method to break the mould was RankTrace, with an unbounded protocol aimed to collect data specifically for preference learning \cite{lopes2017ranktrace}. Nevertheless, GTrace and its derivatives remain one of the most popular annotation tools to this day. 

The core strength of one-dimensional labelling tools is the reduced cognitive load they cause to annotators compared to multi-dimensional labelling in which annotators are requested to split their attention \cite{cowie2013gtrace} across multiple dimensions and labels. Focusing on one affect dimension at a time reduces noise and provides a higher face validity. Multi-dimensional tools, however, can produce annotations at a higher rate. Moreover, as labels across different dimensions are collected simultaneously, the labels are less susceptible to recency effects \cite{erk2003emotional} compared to those obtained from repeated one-dimensional protocols. For all aforementioned reasons, multi-dimensional annotation methods are still popular today. Meanwhile a series of new such tools have being developed recently including DARMA \cite{girard2018darma} (released in 2018) and two versions of RCEA \cite{zhang2020rcea,xue2021rceavr} (the original was released in 2020 and the VR version was released in 2021). 

Another, even more recent, trend is the shift from annotation tools which are usable in a lab setting towards those that are usable in the wild. Many traditional and popular tools require researcher oversight, which limits the dataset size collected. In addition to this issue, the new social reality brought on by the COVID-19 pandemic pushed the field even more towards crowd-sourcing affect labels. PAGAN \cite{melhart2019pagan} is one of the first frameworks developed with crowd-sourcing capacities in mind. Compared to earlier annotation tools, PAGAN is highly configurable and supports multiple different annotation techniques including a GTrace and a RankTrace variant. While PAGAN tackles the issue of crowd-sourcing by using an online platform, other annotation tools offer mobile integration (RCEA \cite{zhang2020rcea}), VR integration (RCEA-360VR \cite{xue2021rceavr}) or aim to alleviate the need for annotators through automatic labelling (NOVA \cite{heimerl2019nova}). We can observe this shift towards crowd-sourcing and mobile integration in custom, yet-to-be-released annotation tools as well \cite{lovei2021designing,salvador2021smartphone}, and we expect more reliable crowd-sourcing methods to appear as the focus will continuously shift from the lab to real-world (in the wild) experiences. 

\section{Game Affective Corpora}\label{sec:corpora}

Instead of building a new affect corpus from scratch one may apply affective computing methods directly to existing corpora. Unlike traditional affective computing datasets, however, player modeling research often focuses on player experience aspects that are not directly linked to affect. Many player experience datasets for instance are annotated with high-level game-related concepts, such as frustration, perceived challenge \cite{yannakakis2010mazeball,karpouzis2015platformer}, engagement \cite{karpouzis2015platformer}, and fun \cite{yannakakis2010mazeball,beaudoin2019funii}. Nevertheless, affective computing methods are still applicable; and conversely, such datasets can offer interesting new test-beds for affective computing applications. This section reviews a number of available affect corpora that are build using games as the underlying context of interaction.

While, traditionally, affective computing datasets are collected through induced emotions and posed expressions, in recent years we witness a notable shift towards spontaneous emotion elicitation and naturalistic settings. During the last decade, a new wave of datasets has started to employ popular multimedia as elicitors of affect \cite{koelstra2012deap,zafeiriou2017affwild,mollahosseini2017affectnet}. Using artefacts such as clips and still images from popular movies has proven to be a reliable and cost-effective way to elicit emotions in a natural way. The resulting datasets, however, only focus on a specific type of passive elicitation that comes with consuming traditional media. The latter half of the decade, instead, has seen a rise in affect corpora that employ interactive elicitation methods including dyads \cite{ringeval2013reloca,kossaifi2019sewa}, group tasks \cite{maman2020game}, board games \cite{doyran2021mumbai}, and videogames \cite{karpouzis2015platformer,beaudoin2019funii,melhart2022again}. 

There are two major differences one can identify between traditional affective computing and game-based corpora. The first difference is that traditional affective computing corpora use third-person annotation (i.e. RELOCA \cite{ringeval2013reloca}, LIRIS-ACCEDE \cite{baveye2015liris}, Aff-Wild \cite{zafeiriou2017affwild}, AffectNet \cite{mollahosseini2017affectnet}, SEWA DB \cite{kossaifi2019sewa}) whereas game-based datasets, instead, use primarily self-reports (i.e. MazeBall \cite{yannakakis2010mazeball}, PED \cite{karpouzis2015platformer}, FUNii \cite{beaudoin2019funii}, MUMBAI \cite{doyran2021mumbai}, AGAIN \cite{melhart2022again}). The second major difference between affective computing and player experience datasets is the wider focus of the latter on experience aspects that often cover behavioural or user states. In fact, most player experience datasets do not consider affect labels or affect manifestations at all. While there is definitely a mapping between affect and higher-level concepts such as fun and engagement, revealing such a relationship might not be as trivial. With this in mind, the survey of datasets presented in this section focuses on game-based datasets that have some connection to affective computing; in particular, they either consider physiological signals and/or are annotated using traditional affective dimensions such as arousal or valence. For the sake of clarity, we also omit datasets that might have been influential in the past but are not available anymore---such as the \emph{Tower Game} \cite{salter2015tower} or the \emph{GeMo} \cite{li2016towards} datasets---and popular game-based datasets that do not include affective labels---such as the \emph{Obstacle Tower} dataset \cite{juliani2019obstacle}. 

\begin{table*}[!tb]
    \centering
    \caption{Affective Game Computing Datasets. ``N/A'' indicates that a category is ``not-applicable''.}
    \resizebox{\textwidth}{!}{\begin{tabular}{|L{2cm}|L{1.5cm}|c|C{1cm}|c|L{2cm}|L{2cm}|L{2cm}|L{1.9cm}|c|}
        \hline
        \textbf{Database} & \textbf{Game Type} & \textbf{Games} & \textbf{Video (hours)} & \textbf{Participants} & \textbf{Modalities} & \textbf{Annotation} & \textbf{Labels} & \textbf{Annotators} & \textbf{Tasks} \\
        \hline
        \hline
        MazeBall \cite{yannakakis2010mazeball} & Navigation & 1 & N/A & 36 & BVP (HRV), EDA, game telemetry & Pairwise & Fun, challenge, frustration, anxiety, boredom, excitement, relaxation & self-report & 1 \\
        \hline
        PED \cite{karpouzis2015platformer} & Platformer & 1 & 6 & 58 & Gaze, head position, game telemetry & Discrete  (5-step), pairwise & Engagement, frustration, challenge  & self-report & 1 \\
        \hline
        GSET Somi \cite{ahmadi2016gset} & Rail Shooter & 1 & 6.75 & 84 & Eyetracking, gameplay video & N/A & N/A & N/A & 1-3 \\
        \hline
        FUNii \cite{beaudoin2019funii} & Action & 2 & N/A & 190 & ECG, EDA, gaze and head position, controller input & Continuous, discrete & Fun (cont.), fun, difficulty, workload, immersion, UX & self-report & 2 \\
        \hline
        RAGA  \cite{granato2020empirical} & Racing (VR~and~PC) & 2 & N/A & 33 & ECG, EDA, EMG, resp. & Continuous bounded & Arousal, valence & self-report & 2 \\
        \hline
        GAME-ON \cite{maman2020game} & Escape room & 1 & 11.5 & 51 & Video, audio, and motion capture data & Discrete (5--9-step) & Emotions, cohesion, warmth, competence, competitivity, leadership, and motivation & self-report & 5 \\
        \hline
        eSports Sensors Dataset \cite{smerdov2020collection} & MOBA & 1 & N/A & 8 & EEG, BVP (HR), EDA, EMG, temp., hand and head gestures & N/A & N/A & N/A & 11 \\
        \hline
        Atari-HEAD \cite{zhang2020atari} & Arcade & 20 & 117 & 4 & Eyetracking, gameplay video, game telemetry & N/A & N/A & N/A & 20 \\
        \hline
        MUMBAI \cite{doyran2021mumbai} & Board-game & 6 & 46 & 58 & Gameplay, facial video, and facial action units  & Discrete labels & Valence, attention, gameplay experience, personality & 56~($3^{rd}$-person) 58~($1^{st}$-person) & 6 \\
        \hline
        BIRAFFE2 \cite{kutt2022biraffe2} & Platformer, navigation & 3 & 23 & 103 & EEG, EDA, game video, game telemetry, facial recognition & Continous (automated) and discrete (survey) & Emotions, NEO-FFI and GEQ survey & automated (emotions), self-report (survey) & 3 \\
        \hline
        AGAIN \cite{melhart2022again} & Racing, shooter, platformer & 9 & 37 & 124 & Game video, game telemetry & Continuous unbounded & Arousal & self-report & 9 \\
        \hline
    \end{tabular}}
    \label{tab:datasets}
\end{table*}

Table \ref{tab:datasets} presents our survey of $11$ datasets 
which can be used for affective game computing research. Most of the datasets presented here are quite recent, (i.e. $8$ of the $11$ are released in 2019 or later) showing the potential and the emerging nature of the field. Moreover, most of the examined datasets focus on one specific context; i.e. the genre of the game. While the AGAIN dataset \cite{melhart2022again} was designed explicitly to offer a wide array of different games, the Atari-HEAD \cite{zhang2020atari} and MUMBAI \cite{doyran2021mumbai} datasets also provide more varied inputs in their own niche (arcade- and board-games, respectively). 
Many game datasets contain video data of the gameplay footage, which is ideal for deep learning applications and for mapping pixels to affect directly \cite{makantasis2019pixels,makantasis2021pixels}. While pixel-based affect detection might be a harder task on datasets collected on board or social games (i.e. MUMBAI \cite{doyran2021mumbai} and GAME-ON \cite{maman2020game}), datasets that provide a large amount of gameplay footage are ideal for computer vision methods (i.e. Atari-HEAD \cite{zhang2020atari} and AGAIN \cite{melhart2022again}). Five out of the eleven affective game computing corpora feature physiological signals that are more commonly used in traditional affective computing research, while only $3$ of them feature eye-tracking data. In addition to traditional features, many game datasets ($5$ of the $11$ surveyed) contain behavioural and contextual data in the form of game telemetry and player input. This type of data has proven to be robust as a predictor of game-related emotional states \cite{melhart2018model,melhart2020moment,melhart2021towards}. 

Even though all datasets surveyed contain multiple modalities of user signals, not all of them offer affect labels. Specifically, $3$ of the surveyed datasets (i.e. GSET Somi \cite{ahmadi2016gset}, eSports Sensors Dataset \cite{smerdov2020collection}, and Atari-HEAD \cite{zhang2020atari}) do not provide affect labels of any sort. The rest of the corpora provide a wide array of experience labels, mostly related to high-level game-related outcomes such as fun, challenge, and engagement. Two of the surveyed datasets feature emotional labels (GAME-ON \cite{maman2020game} and IRAFFE2 \cite{kutt2022biraffe2}) whereas $3$ of them contain at least one type of affective label (RAGA \cite{granato2020empirical}, MUMBAI \cite{doyran2021mumbai}, and AGAIN \cite{melhart2022again}). As mentioned earlier, affect labels in games are predominately self-reported and not usually annotated by a third person. 

Reflecting over Table \ref{tab:datasets} one can observe a substantial increase in interest and availability of game-based datasets over the last few years; there are still, however, many aspects of player experience that have not made it to an affect corpus as of yet. Moreover the design of the various data collection protocols comes with limitations. Most notably the wide spectrum of labels used across the corpora makes it hard to compare datasets and transfer any knowledge gained across corpora. While it is understandable that game-related outcomes are more important than basic affective dimensions for game user researchers, some level of standardisation would go a long way towards making these datasets more accessible and reusable. Nevertheless, affective corpora---as the ones surveyed in Table \ref{tab:datasets}---that are based on interactive elicitation bring a new perspective to affective computing at large, as such elicitation methods were largely unsupported by traditional datasets up until recently.

\section{Considerations and Discussion}\label{sec:discussion}

After surveying the affective game computing field as a whole, the available tools for data collection and finally the datasets available, in this section we will discuss a number of considerations that are linked to this field. Specifically, we will start by outlining a number of ethical considerations and we will then briefly touch upon the current hardware limitations that can be detrimental to research advancements and breakthroughs in the field.

\subsection{Ethical Challenges}

With the acceleration of widespread AI adoption, various ethical considerations around the field have become more important than ever. While, on the one side, we can observe an unprecedented increase of AI use in both academia and industry, on the other side, we can see the growing public anxiety towards the very same systems \cite{bryson2011just,yu2018building}. Although many ethical frameworks exist to address these anxieties, the industry as a whole has been slow to react \cite{hauselmann2021fit}. This section highlights some of the challenges in \emph{Responsibility}, \emph{Transparency}, \emph{Auditability}, \emph{Incorruptibility}, and \emph{Predictability} \cite{bostrom2014ethics} that future game affective applications will have to face \cite{mikkelsen2017ethical,vakkuri2019ethically}. The interested reader may refer to \cite{melhart2023ethics} for a recent in-depth discussion of ethical considerations related to the various uses of AI in and for games. 

The first challenge is to establish a well-defined chain of \emph{responsibility} and ownership over affect models, the data they train on, and their output. A major issue in this topic is the decoupling of the data, the model, and its output. Because the results provided by an affect model are thought to be ``inferred'' instead of ``observed'', the chain of responsibility becomes opaque \cite{wachter2019right}. Even though the responsibility of transparent data handling is well-defined in documents such as European Union's General Data Protection Regulation (GDPR) \cite{voigt2017eu}, the research community is still lagging behind \cite{jost2020two}. The issue is not trivial both on the academic and industrial levels by the fact that affective corpora---despite being deeply personal---is not protected under the current legal frameworks \cite{hauselmann2021fit}. We already note proposals for a change in regulations \cite{siegmann2022brussels} addressing some of these issues, and expect future affective interaction applications to be upheld to more scrutiny.

\emph{Transparency} is a core challenge of AI as a whole. While seemingly a clear-cut issue which can be tackled by explainable AI systems \cite{das2020opportunities,zhu2021open}, in reality, the issue is complicated by the \emph{transparency–efficiency trade-off} \cite{ishowo2019behavioural}. This phenomenon describes the detrimental effect of bias against AI \cite{rovatsos2019we} to human-machine cooperation. Resolving transparency in games is even more challenging due to the ``smoke and mirrors'' nature of the medium.

\emph{Auditability} is more of an industry-specific challenge rather than an academic one, as recent research studies already strive for external validity and reliability. Industrial applications of affective game computing should consider the audit process during their development cycle. As the field moves more and more towards large-scale \emph{foundation models} \cite{bommasani2021opportunities}, the role of planned audits to maintain \emph{transparency} will become more and more a necessity.

The challenge of \emph{Incorruptibility} refers to the robustness of the system against any kind of manipulation. While short-lived research-focused projects are generally not expected to receive many adversarial attacks---if any; industry applications and even large-scale crowd-sourced studies have to face external attackers. However, beyond malicious users, computer models can also be corrupted by internal forces. Biased affect models can encode and perpetuate socioeconomic and sociopolitical biases that lead to direct or indirect harm to the users \cite{yapo2018ethical,gebru2020race,gebru2021datasheets}. Unfortunately, \emph{incorruptibility} is a wicked problem in the field as often there is no apparent way of ascertaining algorithmic bias before the system is deployed.

The final challenge for ethical AC applications is \emph{Predictability}. Similarly to \emph{auditability}, academic studies in general fare well on this front because \emph{predictability} is somewhat analogous to the internal reliability and validity of the system. Beyond this, \emph{predictability} can help increase \emph{transparency} and \emph{incorruptibility} by ensuring the reliability and fairness of the application in question \cite{crowley2019toward}.

The framework presented here as adapted from \cite{melhart2023ethics} describes the universally understood challenges when it comes to AC applications in games. These issues have served as a bedrock for discussions and proposals surrounding ethical player modelling \cite{mikkelsen2017ethical}, ethically aligned design \cite{ieee2019ethics}, ethics in human-AI interaction \cite{crowley2019toward}, AI trustworthiness \cite{thiebes2021trustworthy}, and general ethical AI guidelines \cite{smuha2019eu}. 

\subsection{Hardware Limitations}

As already mentioned in Section \ref{sec:dataCollection} sensing affect via objective measurements offers rich information about the player’s experience; a major limitation, however, is that several of these sensors can be invasive, impractical or even impossible in the wild. \emph{Pupillometry} and \emph{gaze tracking} for instance, are sensitive to variations in light and screen luminance. Camera-based sensing (e.g. facial expressions and body posture) requires a well-lit environment which is often not available when we play videogames in our living rooms for instance. However, the recent rapid advancement of mobile, smartwatch and VR hardware with integrated eye tracking and physiology sensors (e.g. see the recent \emph{HP Omnicept} headset gives such sensing technologies entirely new opportunities and use within games \cite{sidorakis2015binocular}. \emph{Speech} and \emph{text} may offer some alternative unobtrusive and highly accessible modalities which are only applicable, however, to games that feature those modalities. This includes games in which a) speech is a control modality \cite{kannetis2009fantasy,yildirim2011detecting}, b) text is used as means of communication across the audience of a streamed game \cite{melhart2020moment}, c) speech or chat is used for multiplayer coordination, d) natural language processing is used as a game control in text-based adventure games or interactive fiction. Finally, existing hardware for EEG, respiration and EMG (if not embedded within a VR headset) requires the placement of sensors on the player's body thereby making those physiological signals intrusive and rather impractical, to say the least. Recent sensor technology advancements however, especially via state-of-the-art VR headsets (see Section \ref{sec:dataCollection}) have revived the use of physiological signals for commercial standard applications \cite{yannakakis2016psychophysiology}. 

\section{The Road Ahead} \label{sec:road}

In this section, we will cover the two areas we think will be the most challenging for future research in \emph{affective game computing}. In particular, we first discuss current and future research in the area of artificial general intelligence (AGI) and emotion in games, and we move on to discuss particular computer vision research areas that appear to be of high value for affective game computing research. We end with a small section dedicated to large language models (LLMs) and their potential impact on affective game computing.

\subsection{AGI and Affective Game Computing}

Damasio's work \cite{damasio1996somatic} suggests that our ability to recognise human emotion across contexts and people of dissimilar moods, cultural backgrounds and personalities, acts as a facilitator of decision-making and general (emotional) intelligence \cite{mayer1993intelligence}. While the research reviewed has reached important milestones, all key findings suggest that any success of affective (game) computing is heavily dependent on the domain, the task at hand, and the context in general. This limitation of \emph{specificity} is also present in games \cite{yannakakis2014emotion}. The vast majority of the studies presented focus on modelling player experience within a particular game and a narrow player base, and under well-controlled conditions (e.g. \cite{yannakakis2011experience,shaker2011game,shaker2013fusing,martinez2014deep} among many). 

Assuming that the game affective loop can be successfully realised within a particular game, the next long-term and ambitious goal for affective game computing is to achieve a good level of generality across games of the same genre of other genres. For affective game computing to become general, models are required to recognise general emotional and cognitive-behavioural patterns across contexts of players and games. So far the literature is rather sparse in this area with only a few available preliminary studies. Early work focused on the ad-hoc design of general metrics of player interest for prey--predator games \cite{yannakakis2005generic,yannakakis2005Ageneric} followed by player experience models that can operate to a satisfactory degree across dissimilar games \cite{LNCS69740267,shaker2015towards,camilleri2017towards}. Recent work in the area is driven by the AGAIN dataset \cite{melhart2021again} through which a number of promising studies have been performed to test for the generality of player arousal models across games \cite{melhart2022again,makantasis2022invariant}. Discovering entirely new representations of player emotive manifestations across games, modalities, and player types appears to be a critical step towards achieving general player affect models. Methods from representation learning and transfer learning could be of great assistance to this cause \cite{makantasis2022invariant}. In the next section, we discuss how such computer vision methods can further advance the study of affective game computing.



\subsection{Computer Vision for Affective Game Computing}

Computer vision research brings methods with great potential for the future of affective game computing. Recent work in the use of convolutional neural networks has shown that it is possible to solely rely on gameplay footage and in-game audio for predicting player affect with high levels of accuracy \cite{makantasis2019pixels,makantasis2021affranknet+,makantasis2021pixels}. What is fascinating about pixel-based affect modelling is that it is general and user-agnostic: it is general as it can eventually represent any affective patterns by observing the game footage pixels; it is user agnostic as it does not rely on any other personal information about players beyond their gameplay (e.g. manifestations of affect via the physiology of facial expressions). These two properties make pixel-based affect models operational in the wild. Making players affect models usable in the wild is key for the models to be usable in games. A recent direction with great promise addressing the operation of affect models in the wild is the use of \emph{privileged information} \cite{makantasis2021privileged}. Privileged information allows models to be trained with game lab data (e.g. including physiology, speech and facial expression) and operate without these modalities which are not available in a player's living room (i.e. in the wild). 

Complementary to the above computer vision methods, self-supervised learning (SSL) methods such as contrastive learning define a recent machine learning paradigm which has been widely and successfully employed for learning general representations of data \cite{le2020contrastive}. SSL methods attempt to learn by processing different views of the same input that have similar representations. Although contrastive
learning is gaining traction for computer vision tasks such methods have found applications in games \cite{trivedi2021contrastive,trivedi2022learning} and affective computing \cite{shen2022contrastive,pinitas2022supervised} only recently. SSL-based affect modelling assumes that affect information is an inherent property of the manifestations of affect and thus can be fused and learned in a contrastive learning manner. Early results of Pinitas et al. \cite{pinitas2022supervised} suggest that when affect is used as a label to contrast multimodal data arousal models achieve supreme classification performance compared to end-to-end classification.

\subsection{LLMs for Affective Game Computing}

Large scale language models such as GPT-3 \cite{floridi2020gpt} and GPT-4 \cite{openai2023gpt4} appear to offer promising, yet largely unexplored, methods for affective game computing. We envision the use of LLMs for learning and interweaving game context and affect demonstrations as e.g. in early experiments reported in \cite{gallotta2023text}. In principle, an LLM that is hosted in a game engine could predict any affective state transition including states where ``the game is more engaging'' and thereby change the surrounding environment or spawn a number of enemies to get the player to a supposedly more engaging state. Learning to generate affect transitions via foundation models builds on the experience-driven procedural content generation paradigm \cite{yannakakis2011experience} but with a contemporary touch. Generative affect-based AI methods including GPT variants and diffusion models \cite{rombach2022high} will likely expand beyond text-to-(affect)-image applications \cite{galanos2021affectgan}, to text-to-video, text-to-3D models and finally text-to-games with prescribed affect patterns \cite{gallotta2023text}; \emph{text} can be replaced (or complemented) by other modalities including images and sound. 

Compared to the very specific game state transitions that an LLM could use to play a game the relationship between game context and player affect appears to be more general \cite{melhart2022again}. One can thus argue that such relationships could potentially be learned from many games rather than one. We could then imagine the future development (or fine tuning) of large foundation models that are capable of representing and inferring affect transitions based on in-game observations and affect demonstrations. Referring to our earlier discussion about AGI, it remains to be seen to which degree we can build a generalised computational game experience designer, and how.  

\section{Conclusions} \label{sec:conclusions}

The domain of games has advanced the ways we represent, model, and annotate emotion over the last decade. It is their highly interactive nature, the complex spatiotemporal and rich behaviour of their users, and the availability of multiple sensor data that have helped such advancements collectively. Affective computing has also advanced game design and development. For instance, affect-driven adaptation and automatic player experience design are becoming gradually the norm across a number of game genres (e.g. horror and racing). This paper surveyed this exciting intersection of AC and games and introduced the field of \emph{affective game computing} through the phases of the affective loop. The paper also introduced a taxonomy of terms and methods used largely in affective game computing and placed exemplar studies on this taxonomy. Finally, this survey offers the interested reader a comprehensive list of data collection tools and datasets that are directly accessible for research in this field.

With this paper we survey the past, capture the present and offer a vision about the future of this field. In our attempt we tried to be as comprehensive and inclusive as possible (as indicated by the reference list of this article). Despite the best of our intentions, however, we are aware that some studies were omitted or not discussed thoroughly due to space considerations. Evidently, while a lot has been achieved in \emph{affective game computing} until nowadays, there are still many open research questions left to be addressed. We hope that this paper will act as a key driver of groundbreaking research and innovation in this emerging field.

\bibliographystyle{IEEEtran}
\bibliography{sample}

\begin{IEEEbiography}[{\includegraphics[width=0.9in,clip,trim={50px 0 100px 0},keepaspectratio]{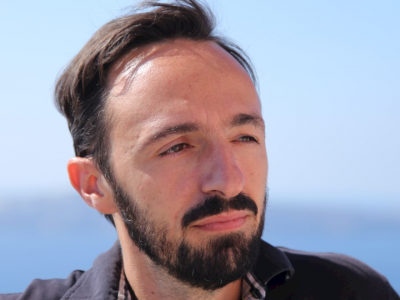}}]
{Georgios N. Yannakakis} is a Professor and Director of the Institute of Digital Games, University of Malta (UM) and a co-founder of modl.ai. He received the PhD degree in Informatics from the University of Edinburgh in 2006. Prior to joining UM, in 2012 he was an Associate Professor at the Center for Computer Games Research at the IT University of Copenhagen. He does research at the crossroads of artificial intelligence, affective computing, games and computational creativity. He has published more than 350 papers in the aforementioned fields and his work has been cited broadly. His research has been supported by numerous national and European grants (including a Marie Skłodowska-Curie Fellowship) and has appeared in \emph{Science Magazine} and \emph{New Scientist} among other venues. He is currently the Editor in Chief of the \emph{IEEE Transactions on Games}, an Associate Editor of the \emph{IEEE Transactions on Evolutionary Computation}, and used to be an Associate Editor of the \emph{IEEE Transactions on Affective Computing} and the {IEEE Transactions on Computational Intelligence and AI in Games} journals. He has been the General Chair of key conferences in the area of game artificial intelligence (IEEE CIG 2010) and games research (FDG 2013, 2020). Among the several rewards he has received for his papers, he is the recipient of the \emph{IEEE Transactions on Affective Computing Most Influential Paper Award} and the \emph{IEEE Transactions on Games Outstanding Paper Award}. He is a senior member of the IEEE.
\end{IEEEbiography}
\vskip 0pt plus -1fil
\begin{IEEEbiography}[{\includegraphics[width=0.9in,clip,trim={75px 0 75px 0},keepaspectratio]{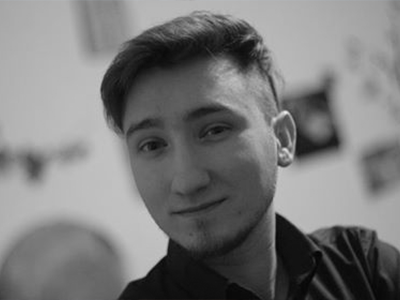}}]
{David Melhart} is a Senior Member of Technical Staff at modl.ai, and a Postdoctoral Researcher at the Institute of Digital Games, University of Malta. He received a MA degree in Cognition and Communication from the University of Copenhagen in 2016 and a PhD degree in Game Research from the University of Malta in 2021. His research focuses on Machine Learning, Affective Computing, and Games User Modelling. He has been the Communication Chair of FDG 2020, a recurring organiser and Publicity Chair of the \emph{Summer School series on Artificial Intelligence and Games} (2018-2023), the Workshop and Panels Chair of FDG 2023, Editorial Assistant to the \emph{IEEE Transactions on Games}, Guest Associate Editor on the \emph{User States in Extended Reality Media Experiences for Entertainment Games} Special Issue of \emph{Frontiers in Virtual Reality and Human Behaviour}, and Review Editor of \emph{Frontiers in Human-Media Interaction}.
\end{IEEEbiography}

\end{document}